\documentclass[10pt,journal,compsoc]{IEEEtran}
\usepackage{amsmath}
\usepackage{amsthm}
\usepackage{amssymb}
\usepackage{graphicx}
\usepackage{multirow}
\usepackage{subcaption}
\usepackage{subfloat}
\usepackage{booktabs}
\usepackage{array}
\usepackage{url}
\usepackage{todonotes}
\usepackage{algorithm}
\usepackage{algorithmic}
\usepackage{setspace}
\usepackage{epstopdf}
\usepackage{booktabs}
\usepackage[neverdecrease]{paralist}
\usepackage[numbers]{natbib}
\usepackage{xspace}
\usepackage{xcolor}
\usepackage{tikz}
\usepackage{tikz-qtree}
\usetikzlibrary{calc, chains, fit,shapes}
\usetikzlibrary{fit}
\usetikzlibrary{matrix,positioning,quotes}
\usetikzlibrary{arrows}
\usepackage{amsmath}
\usepackage{amssymb}
\usepackage{amsthm}
\usepackage{dblfloatfix}
\pdfoutput=1

\ifCLASSOPTIONcompsoc
\else
\fi
\ifCLASSINFOpdf
\else
\fi
\begin{document}
%
\title{ExpFinder: An Ensemble Expert Finding Model Integrating $N$-gram Vector Space Model and \hits}
%
%
%

\author{Yong-Bin~Kang,
        Hung Du,
        Abdur Rahim Mohammad Forkan,
        Prem Prakash Jayaraman,
        Amir Aryani,
        Timos Sellis,~\IEEEmembership{Fellow,~IEEE}

\IEEEcompsocitemizethanks{\IEEEcompsocthanksitem Yong-Bin Kang is with Department of Media and Communication, Swinburne University of Technology, Australia \protect\\
\textbf{E-mail:} ykang@swin.edu.au

\IEEEcompsocthanksitem Hung Du, Abdur Rahim Mohammad Forkan, Prem Prakash Jayaraman, and Timos Sellis are with Department of Computer Science and Software Engineering, Swinburne University of Technology, Australia \protect\\
\textbf{E-mail:} hungdu@swin.edu.au, fforkan@swin.edu.au, pjayaraman@swin.edu.au, and tsellis@swin.edu.au

\IEEEcompsocthanksitem Amir Aryani is with Social Innovation Research Institute, Faculty of Business and Law, Swinburne University of Technology, Australia. 
\textbf{E-mail:} aaryani@swin.edu.au
}

\thanks{}}

\newcommand{\mc}{\mathcal}
\newcommand\var{\texttt}
\newcommand\tb{\textbf}
\newcommand{\mb}{\mathbf}
\newcommand{\mn}{\mathrm}
\newcommand{\lm}{\fontfamily{lmss}}
\newcommand{\EF}{{ExpFinder}\xspace}
\newcommand{\ecg}{{ECG}\xspace}
\newcommand{\hits}{{$\mu$CO-HITS}\xspace}
\newcommand{\yb}[1]{\textcolor{red}{\small\textbf{[YB]}#1$\triangleleft$}}
\newcommand\scalemath[2]{\scalebox{#1}{\mbox{\ensuremath{\displaystyle #2}}}}
\renewcommand{\IEEEbibitemsep}{0pt plus 2pt}
\makeatletter
\IEEEtriggercmd{\reset@font\normalfont\footnotesize}
\makeatother
\IEEEtriggeratref{1}

\newcommand{\peter}[1]{\textcolor{green}{\small\textbf{[Peter]}#1$\triangleleft$}}
\newcommand{\prem}[1]{\textcolor{blue}{\small\textbf{[Prem]}#1$\triangleleft$}}
\newcommand{\forkan}[1]{\textcolor{orange}{\small\textbf{[Forkan]}#1$\triangleleft$}}
\newcommand{\amir}[1]{\textcolor{red}{\small\textbf{[Amir]}#1$\triangleleft$}}
\newcommand{\timos}[1]{\textcolor{red}{\small\textbf{[Timos]}#1$\triangleleft$}}

\IEEEcompsoctitleabstractindextext{%
\begin{abstract}
Finding an expert plays a crucial role in driving successful collaborations and speeding up high-quality research development and innovations. However, the rapid growth of scientific publications and digital expertise data makes identifying the right experts a challenging problem. Existing approaches for finding experts given a topic can be categorised into information retrieval techniques based on vector space models, document language models, and graph-based models. In this paper, we propose \textit{\EF}, a new ensemble model for expert finding, that integrates a novel $N$-gram vector space model, denoted as $n$VSM, and a graph-based model, denoted as \textit{\hits}, that is a proposed variation of the CO-HITS algorithm. The key of $n$VSM is to exploit recent inverse document frequency weighting method for $N$-gram words, and \EF incorporates $n$VSM into \hits to achieve expert finding. We comprehensively evaluate \EF on four different datasets from the academic domains in comparison with six different expert finding models. The evaluation results show that \EF is an highly effective model for expert finding, substantially outperforming all the compared models in 19\% to 160.2\%. 
\end{abstract}
}


\maketitle


%
\IEEEpeerreviewmaketitle

\section{Introduction}\label{sec:intro}
{Finding experts in a particular domain} is key to accelerate rapid formation of teams to respond to new  opportunities, as well as undertake and address new frontiers in research innovations. Further, accurately identified experts can significantly contribute to enhancing the research capabilities of an organisation leading to higher quality research outcomes. In general, an \textit{expert} is defined as a person who has sufficient knowledge and skills in a given field~\citep{husain2019expert}. Such knowledge and skills are called \textit{expertise}. While digitally available data  (e.g. scientific publications) describing expertise of experts is rapidly growing, manually collating such information to find experts seems impractical and expensive. Thus, often in a large research organisation with diverse disciplines, finding experts in a field that one does not know or has limited knowledge is particularly very challenging.

Information retrieval techniques have been widely used to aid retrieval task for finding experts from digitally available expertise data (we collectively term these as \textit{documents} in this paper)
 such as scientific publications~\citep{Stankovic2010LookingFE}. 
Based on the literature~\cite{gonccalves2019automated}, there are two specific tasks for expert retrieval: (1) \textit{expert finding} - identifying experts given a topic from available documents and rank them based on their expertise level, and (2) \textit{expert profiling} - identifying the areas of expertise given an expert.
In this paper we focus on the first task (i.e. {expert finding}) and propose an ensemble model for it from unstructured documents. We use the term \textit{topic} to represent a field of  expertise.


Most existing approaches for expert finding are based on \textit{vector space models} (VSM), \textit{document language models} (DLM), or \textit{graph-based models} (GM).
In VSM, expert finding is often solved by modeling the weights of topics, associated with the documents produced by experts, using Term Frequency-Inverse Document Frequency (TFIDF) or its variation~\citep{chuang2014combining,alhabashneh2017fuzzy}. 
In DLM, expert finding is achieved by estimating the probability that a topic would be observed in the documents of an expert~\citep{Balog:2009, Wang:2015,WISER:2019}. 
In GM, a graph is used to represent associations among experts, documents and/or topics. The  strengths of the associations are inferred to estimate the expertise degree of an expert given a topic using various graph analytics such as expert-document-term association paths~\citep{Gollapalli:2013}, Hyper-Induced Topic Search (HITS)~\citep{Campbell:2003,Yeniterzi:2014}, or social network based link analysis methods~\cite{faisal2019expert,bok2019expert,Sziklai:2018}.


Although various models in these approaches aforementioned have been proposed, integrating VSM and GM for expert finding has been little studied in the literature. In this work, we propose an ensemble model for expert finding, \textit{\EF}\footnote{ExpFinder's source code is publicly available on\\ \url{https://github.com/Yongbinkang/ExpFinder}}, that integrates a novel $N$-gram VSM, denoted as $n$VSM, with a GM using an \textit{expert collaboration graph (ECG)}. We develop $n$VSM for estimating the expertise degree (or \textit{weight}) of an expert given a topic by leveraging the recent Inverse Document Frequency (IDF) weighting~\citep{nidf:2017} for $N$-gram words (simply $N$-grams) composed of two or more terms (for N$>$1). This method demonstrated a higher robustness and effectiveness in measuring the IDF weights of $N$-grams.
We also build an \ecg in the form of an expert-document bipartite graph to represent the associations between experts and documents based on the co-authorship information.
To estimate the weight of an expert given a topic on the \ecg, we propose the GM, \hits, that is formed by applying two variation schemes to the generalised CO-HITS~\citep{Hongbo:2009} algorithm. 


This paper makes three main contributions. First, to our best knowledge, \EF is the first attempt to introduce $n$VSM for expert finding. 
Second, we propose \EF an ensemble model that combines $n$VSM and the GM using \hits to create a stronger model for expert finding that achieves better performance than a single one. \EF incorporates the weights of experts estimated by $n$VSM into an \ecg and uses \hits on the \ecg to better estimate the weights of experts for a given topic. Third, we conduct comprehensive empirical evaluations to measure the effectiveness of \EF using four different datasets (LExR~\cite{Vitor:2016} and three DBLP datasets~\cite{bordea2013benchmarking}) in academic domains and compare the results with six different expert finding models: the  TFIDF-based VSM, two DLMs~\citep{Balog:2009,WISER:2019} and three GMs~\citep{Gollapalli:2013,Daniel:2015,Hongbo:2009}. 


This rest of the paper is organised as follows. Section~\ref{sec:related_work} provides related works in expert finding. Section~\ref{sec:overview} presents an overview of \EF and Section~\ref{sec:design} discusses in-depth steps for building \EF. 
Section~\ref{sec:eval} presents thorough empirical evaluations of \EF, followed by conclusion in Section~\ref{sec:conclusion}.

\section{Related Work}\label{sec:related_work}

In recent years, with the growing amount of digital expertise sources, expert finding has become an intensive research area in information retrieval community~\cite{gonccalves2019automated}.
We can mainly classify expert finding approaches into three categories: VSM, DLM and GM. 

In the VSM approach,
the common idea is to estimate relevance between a document and a topic using a weighting scheme in VSM (e.g. TFIDF or its variation). Then, finding experts can be done by assuming that an expert is seen as the collection of its published documents $\mc{D}_x$. That is, the weight of an expert $x$ given a topic $t$ is estimated by aggregating relevance scores between each document in $\mc{D}_x$ and $t$.
For example,  TFIDF  was used to find experts in community question answering websites in which the goal is to find users with relevant expertise to provide answers for given questions~\cite{riahi2012finding}. A variation of TFIDF was also applied for expert finding in an organization’s ERP system~\cite{Schunk:2010}. The work~\cite{chuang2014combining} also used TFIDF to identify experts given a topic using a topic extension method (finding interrelated terms of a given topic from the corpus), where TFIDF was used to estimate relevance between extended terms and each expert's documents. TFIDF was also used to estimate the weights of topics indicating the interests of an expert, and this information is used with fuzzy logics for expert finding~\cite{alhabashneh2017fuzzy}.



The aim of the DLM approach
is to find experts whose documents are directly related to a given topic.
In common, this approach estimates the relationships between a topic and an expert as the probability of generating the topic by the expert \cite{Balog:2009}, or between an expert and its publications~\cite{Mangaravite:2016}. BMExpert~\cite{Wang:2015} used the DLM~\cite{Balog:2009} for expert finding using three factors: relevance of documents to the topic, importance of documents, and associations between documents and experts. Similarly, the work \cite{van2016unsupervised} used a probabilistic DLM for expert finding by probabilistically generating a textual representation of an expert according to his documents and then ranking such documents according to a given topic. Recently, a probabilistic model, \textit{WISER}~\cite{WISER:2019}, estimated the importance of experts' documents given a topic using  BM25~\cite{BM25:2009}. Using this importance, such documents were ranked and these ranks were summed to represent the topic-sensitive weight of an expert.


In the GM approach,
experts are represented as nodes, and their relationships are represented by their edges or implicitly derived from a  graph. Different algorithms were used in the GM approach, such as Hyperlink-Induced Topic Search (HITS)  \cite{Campbell:2003, Yeniterzi:2014, jiang2016exploiting} and PageRank \cite{koumenides2014ranking}. 
For expert finding, PageRank was adapted in the context of online community discussions on a user-user graph built based on votes from users whose questions were answered by whom~\cite{zhang2007expertise}. Also, a modified PageRank algorithm was developed and applied for finding experts in online knowledge communities~\cite{wang2013expertrank}.
HITS is also a graph-based link analysis algorithm originally designed for ranking the importance of web pages based on authority and hub scores. The work~\cite{Campbell:2003} built an expert-expert bipartite graph based on email communication patterns and attempted to find the ranking of experts using HITS. 
CO-HITS was introduced~\cite{Hongbo:2009} to incorporate a bipartite graph with the content information from both sides (e.g. experts and documents in our context) by adding personalised parameters to HITS, and CO-HITS showed higher performance than HITS~\cite{Hongbo:2009}. Using an author-document-topic (ADT) graph, the expert finding GM model~\cite{Gollapalli:2013} leveraged possible paths between a topic and an expert on the ADT graph. Recently, diverse expert finding approaches were proposed in a social network. For example, the authors \cite{faisal2019expert} proposed a method for finding experts who can answer questions in a social network for `community question answering' using users' votes and reputations. The approach \cite{bok2019expert} focused on finding experts who can answer users’ questions based on 
users' online social activities in a social network (e.g. Twitter). 

Also, we observe that some models tend to mix different techniques among DLM, VSM, and/or GM  \cite{kundu2019formulation}. For example,
AuthorRank \cite{deng2011enhanced} combined a generative probabilistic DLM and a PageRank-like GM based on community engagement of expert candidates. The DLM was used to identify the most relevant documents, while the GM was used to model the authors’ authorities based on the community co-authorship.
The work~\cite{liu2013integrating} combined a cluster-based language model and a VSM for finding experts in question and answer communities. The authors \cite{kundu2019formulation} proposed a complex model for community question answering  using a variation of the DLM~\cite{Balog:2009} and a HITS-based GM (the HITS algorithm on a competition based expertise network~\cite{Aslay:2013}), where the scores from these models were linearly combined to rank experts given a question. 
The work~\cite{Mahani:2018} used the Dempster-Shafer combination theory to combine the DLM~\cite{Balog:2009} and a graph algorithm that analyses a social interaction of experts. However, this work did not provide technical details on how such combination is done.

Differing from the above approaches, \EF 
is a first attempt in devising an ensemble model that incorporates $n$VSM  into \hits on an \ecg. The proposed   $n$VSM takes advantage of the IDF weighting for $N$-grams~\cite{nidf:2017}. \hits is a novel variation of CO-HITS and runs on an \ecg (i.e. expert-document bipartite graph), compared to previous works~\citep{Campbell:2003,Yeniterzi:2014}  that applied HITS on an expert-expert graph.

\section{Introduction to ExpFinder}\label{sec:overview}
In this section, we present the overview of \EF, and  the basic notations that we will use in the paper. 

\subsection{Overview of ExpFinder}
\EF aims to identify ranked experts according to their expertise degree given a topic. In this paper, we assume that a topic is represented as a \textit{noun phrase} which is extracted from documents (e.g. scientific publications) of experts in a given domain. The reason is that domain-specific concepts are often described by noun phrases that represent the key information within a given corpus~\cite{Kang:2014}. 
A noun phrase  means  a  single-word noun or a group of words that function together as a  noun.  

The key of \EF is the utilisation of \textit{two knowledge facets} in a unified manner. The one is the estimation of the weights of experts given a topic by utilising information in the proposed $n$VSM. 
The second facet is \hits 
that performs on an expert collaboration graph (\ecg), where the expert collaboration is measured by the joint production of experts (e.g. co-authored documents). We incorporate the result of $n$VSM into the \ecg, and reinforce the weights of experts given a topic using \hits.
The following presents the key steps in \EF (see also Fig.~\ref{fig:system_overview}):


\begin{figure}[!t]
\centering
\includegraphics[trim=0cm 0cm 0cm 0cm, clip,width=240pt]{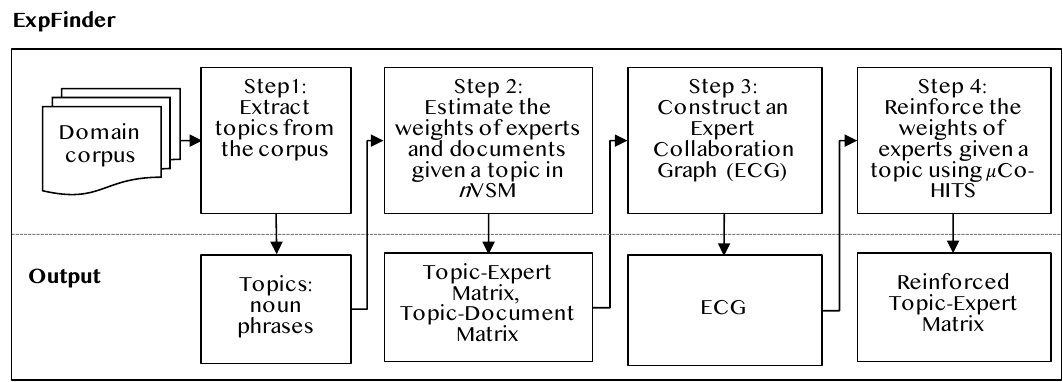}
\caption{The overview of \EF.}\label{fig:system_overview}
\vspace{-10pt}
\end{figure}

{\textbf{Step 1: Extract topics}}: Given experts and their documents (also called \textit{corpus}) in a given domain, we extract noun phrases as topics. 
    
    
\textbf{Step 2: Estimate the weights of experts and documents given topics}: Given a topic, we estimate the weights of experts and documents based on the proposed \textit{$n$TFIDF} method in $n$VSM. In this paper, we also call such weights  \textit{topic-sensitive weights} as these weights are sensitive to the given topic. 
Given a topic, the key of $n$TFIDF lies in a combination of the frequency of the topic with the IDF method of $N$-grams over the corpus~\citep{nidf:2017}. The  output of this step includes a topic-expert matrix and a topic-document matrix, where an entry reflects the weight of an expert and a document given a topic, respectively.
    
\textbf{Step 3: Construct an ECG}: We construct an \ecg to represent associations between experts and their jointly-published documents. This graph is modelled by a \textit{directed, weighted bipartite graph} that has two kinds of nodes, one representing experts and the other representing documents. 
A directed edge points from a document $d$ to an expert $x$, if $x$ has published $d$. 

\textbf{Step 4: Reinforce expert weights using \hits}:
As presented above, to rank experts, \EF integrates the two knowledge facets: (1) $n$VSM to estimate the weights of the experts and documents given a topic ({\textbf{Step 2}}); and (2) \hits incorporating such weights into an \ecg ({\textbf{Step 3}}) to further reinforce the weights of experts. 
The outcome of this step is the reinforced topic-expert matrix showing the weights of experts. Finally, we rank the experts for each topic from the matrix.

\subsection{Notations}\label{sec:notation}
We present the following basic notations in this paper.
\begin{itemize}
    \item Let $\mc{X}$ be the set of experts, and $|\mc{X}|$ be the number of experts in $\mc{X}$.
    \item Let $\mc{D}$ be the set of all documents published by $\mc{X}$. Let $\mc{D}_x$ be all documents published by $x \in \mc{X}$. Also, let $\mc{X}_{d}$ denotes the set of the experts that have a document $d \in \mc{D}$
    \item Let $\mc{T}$ be the set of topics extracted from $\mc{D}$. 
    
    \item Let $\mb{TX}$ be a $|\mc{T}| \times |\mc{X}|$ topic-expert matrix where rows and columns are labeled with $\mc{T}$ and $\mc{X}$, respectively. The entry that lies in the $i$-th row and the $j$-th column of $\mb{TX}$ is denoted as $\mb{TX}_{i,j}$ that indicates the weight of $x_j \in \mc{X}$ on $t_i \in \mc{T}$. 
    If a weight is higher, the more important the corresponding expert is on the given topic. 

    \item Let $\mb{DX}$ be a $|\mc{D}| \times |\mc{X}|$ document-expert matrix where rows and columns are labeled with $\mc{D}$ and $\mc{X}$, respectively. The entry of $\mb{DX}_{i,j}$ shows the weight of an expert $x_j \in \mc{X}$ on a document $d_i \in \mc{D}$ based on $x_j$'s contribution towards $d_i$.
    
    
    \item Let $\mb{TD}$ be a $|\mc{T}| \times |\mc{D}|$  topic-document  matrix where rows and columns are labeled with $\mc{T}$ and $\mc{D}$, respectively. $\mb{TD}_{i,j}$ represents the weight of document $d_j \in \mc{D}$ on $t_i \in \mc{T}$. 
   If a weight is higher, the more important the corresponding document is on the given topic. 

\end{itemize}

\section{Design of ExpFinder}\label{sec:design}
In this section, we present the details of the four steps for designing and developing \EF.

\subsection{Extract Topics} \label{sec: keyphrase}
As presented in Section~\ref{sec:overview}. we assume that a topic is represented as a {noun phrase}. 
We perform the following steps to extract noun phrases from $\mc{D}$. First, for each document $d \in \mc{D}$, we split $d$ into its sentences keeping their sequential indices.
Second, for each sentence, we analyse POS tags of the words in the sentence and remove stopwords. POS tagging is the process for assigning a part of speech to each word in a sentence. Then, each word remained is converted into its lemmatised form. Lemmatisation is the process of grouping together the inflected forms of a word, thus they can be considered to be a single item (e.g. `patients' is lemmatised to `patient'). Third, in the sentence, we use the following linguistic pattern based on POS tags to extract noun phrases: 
\begin{equation}\label{eq:keyphrase}
{\rm (JJ)^*|(VBN)^*|(VBG)^*(N)^+},     
\end{equation}
where `${\rm JJ}$' means adjective, `${\rm VBN}$' past participle, `${\rm VBG}$' gerund, and `${\rm N}$' nouns. Using this pattern, we can extract a noun phrase starting with (1) one or more nouns; (2) one or more adjectives followed by one or more nouns (e.g. `medical system'); (3) one or more past participle followed by one or more nouns (e.g. `embedded system'); and (4) one or more gerund followed by one or more nouns (e.g. `learning system'). The symbol `*' denotes zero or more occurrences, `+' denotes one or more occurrences.


Note that \EF does not rely on a particular method for noun phrase extraction, and thus can incorporate any noun phrase extraction methods.

\subsection{Estimate the weights of experts and documents given topics}\label{sec:ntfidf}
We now present the process for creating a topic-expert matrix $\mb{TX}$ and a topic-document matrix $\mb{TD}$ from the extracted topics using $n$TFIDF in $n$VSM. These matrices will be used as the input to \hits.

\subsubsection{Topic-Expert Matrix Creation}\label{sec:XT}
To create a $\mb{TX}$, our fundamental is to utilise the definition of the DLM~\cite{Balog:2009,Wang:2015} for expert finding. Thus, we first briefly describe how this DLM can measure the topic-sensitive weight of an expert $x \in \mc{X}$ given a topic $t \in \mc{T}$, denoted as $p(x|t)$. Formally, it is given as~\citep{Balog:2009}:
\begin{equation}\label{eq:pw}
    p(x|t) = {p(x, t)}/{p(t)},
\end{equation}
where $p(x, t)$ is the joint probability of $x$ and $t$, and $p(t)$ is the probability of $t$. We can ignore $p(t)$ as this is a consistently constant over all experts $\mc{X}$. Thus, $p(x|t)$ is approximated by $p(x,t)$ that is reformulated  considering documents $\mc{D}_x$~\cite{Balog:2009}:
\begin{equation}\label{eq:pw_2}
    \begin{split}
    p(x, t) &= \sum_{d \in \mc{D}_{x}}p(x, d, t) = \sum_{d \in \mc{D}_{x}}p(d)p(x, t|d)  \\
    &= \sum_{d \in \mc{D}_{x}}p(d)p(t|d)p(x|d).
    \end{split}
\end{equation}

In Eq.~\ref{eq:pw_2}, we observe the following notations~\citep{Wang:2015}:
\begin{itemize}
    \item $p(d)$ is the prior probability of $d$ that can also be interpreted as the weight (or importance) of $d$.

    \item $p(x|d)$ is the conditional probability of $x$ given $d$ (e.g. in a simply way, it can be estimated based on the order of $x$ in the co-author list in $d$~\cite{Wang:2015}). 
    \item $p(t|d)$ is the conditional probability of $t$ given $d$. 
\end{itemize}

In the DLM~\citep{Balog:2009,Wang:2015},  it is assumed that a document $d$ is described as a collection of terms that appear in $d$. An importance of a term $w \in t$ within $d$ is determined by the proportion of its occurrences. DLMs provide a way of capturing this notion by representing a document as multinomial probability distribution over the vocabulary of terms. 
To estimate $p(t|d)$, let $\theta_d$ be the document model of $d$, and the probability of $t$ in $\theta_d$ is $p(t|\theta_d)$. This $p(t|\theta_d)$ indicates how likely we see $t$ if we sampled $t$ randomly from $d$. Thus, $p(t|d)$ is rewritten as $p(t|\theta_d)$ taking the product of $t$'s individual term probabilities as follows~\cite{Balog:2009,Wang:2015}:  
\begin{equation}\label{eq:topic_by_terms}
    p(t|\theta_d) = \prod_{w\in t} p(w|\theta_d),
\end{equation}
where $w$ is an individual term in $t$. However, a limitation of Eq.~\ref{eq:topic_by_terms} is that unseen terms in $d$ would get a zero probability. Thus, it is a common in DLMs to introduce a \textit{smoothing} factor to assign non-zero probability to the unseen terms.
Typically, it can be done by reducing the probabilities of the terms seen in the corpus and assigning the additional probability mass to unseen terms. Formally, $p(w|\theta_d)$ is re-expressed as:
\begin{equation}\label{eq:document_model}
    p(w|\theta_d) = (1 - \lambda_\theta) p(w|d) + \lambda_\theta p(w|\mc{D})
\end{equation}
where $p(w|d)$ is estimated by the term frequency of $w$ in $d$ divided by $|d|$ (the number of terms in $d$), denoted as $tf(w,d)$, and $p(w|\mc{D})$ is the term frequency of $w$ in $\mc{D}$ normalised by $|\mc{D}|$, i.e., ${tf(w, |\mc{D}|)}$. The parameter $\lambda_\theta$ controls the influence of the two probabilities.

We now present our novelty for estimating $p(t|d)$ using $n$TFIDF. Since $n$TFIDF is an extension of TFIDF, we briefly describe how $p(t|d)$ can be estimated using TFIDF in VSM.
In a sense, $p(t|d)$ can also be interpreted using TFIDF~\citep{Roelleke:08}. 
Note that TFIDF is a measure based on the distance between two probability distributions, expressed as the cross-entropy: (1) a \textit{local distribution} of $w \in t$ in $d$, and (2) a \textit{global distribution} of $w$ in $\mc{D}$. 
TFIDF is a measure of perplexity between these two distributions. 
A higher perplexity score implies a higher relevance of $d$ to $w$. The cross-entropy between distributions $p_w$ and $q_w$ is as follows:

\begin{equation}\label{eq:entropy}
    -\sum_w p_w \log q_w= \sum_w p_w \log \frac{1}{q_w},    
\end{equation}
if we substitute $p_w$ with ${tf(w,d)}$ (TF) and $\frac{1}{q_i}$ with the inverted probability of encountering $d$ with a term $w$ (IDF), denoted as $\frac{|D|}{df(w)}$, where $df(w)$ is the document frequency of $w$, we obtain a TFIDF formula:

\begin{equation}\label{eq:entropy2}
    p(t|d) \approx \sum_{w\in t} {tf(w,d)} \log \frac{|D|}{df(w)}.
\end{equation}
Thus, as highlighted in~\cite{lu2013insight}, VSM and DLM are actually closely related. The TF component ${tf(w,d)}$ is exactly same as the probability of seeing a term $w$ in DLM.
The IDF component $\frac{|D|}{df(w)}$ is implicitly related to a smoothing method in DLM that uses the collection frequency (${tf(w, |\mc{D}|)}$: term frequency of $w$ in $\mc{D}$ normalised by $|\mc{D}|$.



Based on the above observation, we now present our approach for estimating $p(t|d)$ using $n$TFIDF in $n$VSM.
Although some variant forms of TFIDF methods have been proposed, the majority of TFIDF methods use the same IDF function~\citep{nidf:2017}.
However, one drawback of IDF is that it cannot handle $N$-grams, contiguous sequence of $N$ terms (for $N$$>$1). The reason is that IDF tends to give a higher weight to a term that occurs in fewer documents. Note that typically, phrases occur in fewer documents when their collocations are less common. Thus, uncommon phrases  (e.g. noise phrases) are unintentionally assigned high weight, yielding the conflict with the definition of a good phrase that constitutes a succinct conceptual descriptor in text. To address it, \textit{$N$-gram IDF} for weighting phrases was recently proposed~\citep{nidf:2017}. $N$-gram IDF has shown the ability to accurately estimate weights of dominant phrases of any length, simply using the domain corpus.

The key in $n$VSM is the proposed formula $n$TFIDF that uses a combination of the frequency of a topic $t$ with $t$'s $N$-gram IDF. As that frequency, we use the average frequencies of the constituent terms in $t$. Formally, using $n$TFIDF, $p(t|d)$ is defined as:

\begin{equation}\label{eq:ntfidf}
\begin{split}
    p(t|d) \approx &~n{\rm TFIDF} (t, d) = ntf(t, d) \cdot nidf(t), {\rm ~where} \\
    &ntf(t, d) = \frac{\sum_{i=1}^{n}{tf ({w_i, d}})}{|t|}, \\
    &nidf(t) =  \log \frac{|\mc{D}|\cdot df(t)}{df(w_1 \land w_2 \land ... \land w_n)^2}
\end{split}
\end{equation}
where $w_1, \ldots, w_n$ are $n$-constituent terms in $t$, 
$tf ({w_i,d})$ is the term frequency of $w_i$ in $d$ normalised by $|d|$, 
$|t|$ is equal to $n$, and $nidf (t)$ is the $N$-gram IDF method for $t$ \citep{nidf:2017}.
Eq.~\ref{eq:ntfidf} applies for all $|t| \geq 1$, where $nidf(t)$ is equal to the log-IDF, $\log \frac{|\mc{D}|}{df(t)}$, in Eq.~\ref{eq:entropy2},  when $|t|$=1.


Finally, in $n$VSM, $p(x|t)$ in Eq.~\ref{eq:pw_2} is calculated using $p(t|d)$ in Eq.~\ref{eq:ntfidf} and is stored into the entry $\mb{TX}_{i(t), i(x)}$, where $i(t)$ and $i(x)$ indicate the row and column index of $t$ and $x$, respectively, in the $\mb{TX}$.



\subsubsection{Topic-Document Matrix Creation}\label{sec:PT}
To create a topic-document matrix $\mb{TD}$, we need to calculate the topic-sensitive weight of a document $d$  given a topic $t$. Following the idea of the DLM~\cite{Balog:2009} again, we can estimate this weight by calculating the probability of $d$ being relevant to $t$: $p(d|t)$. Using the Bayes theorem, $p(d|t)$ can be calculated as:

\begin{equation}\label{eq:pw_d}
    p(d|t) = {p(t|d) p(d)}/{p(t)},
\end{equation}
where  $p(t)$ can be ignored as it is a consistently constant over all documents. Thus, $p(t|d)p(d)$ can be calculated by multiplying  $p(t|d)$ in Eq.~\ref{eq:ntfidf} and $p(d)$.
Finally, $p(d|t)$ is stored into the entry $\mb{TD}_{i(t), i(d)}$, where $i(t)$ and $i(d)$ indicate the row and column index of $t$ and $d$, respectively, in the $\mb{TD}$.

\subsection{Construct an ECG}
Although we have estimated the topic-sensitive weight of an expert $x$ for a given topic $t$ in $n$VSM, one potential limit may be that $p(x|t)$ in Eq.~\ref{eq:pw_2} mainly relies on the documents $\mc{D}_x$ (i.e. $\sum_{d \in \mc{D}_{x}}$), ignoring the social importance (or influence) of experts.
Our premise is that the expertise degree of $x$ on $t$ can depend not only on $x$'s knowledge on $t$, but also on $x$'s social importance among $t$'s collaborating experts in a given domain. Thus, we propose that an expert collaboration graph (i.e. \ecg) can also be a valuable source, in order to estimate such social importance. This estimation is achieved by identifying more \textit{authoritative} (or influential) {topic-sensitive experts} considering their joint documents. That is, in a sense, an \ecg is a social network for experts,
and \EF calculates the authority score of $x$ by repeatedly exploring the collective importance of the joint documents, published by $x$ and $x$'s coauthors, using \hits over the \ecg. More specifically, \EF incorporates the topic-sensitive weights of experts given topic, estimated by $n$VSM, into an \ecg and reinforces such weights using \hits.

Let ${G} = (V, E)$ be an \ecg (i.e. directed, weighted bipartite graph) that has two node types: experts $\mc{X}$ (also called \textit{authorities}) and documents $\mc{D}$ (also called \textit{hubs}). Thus, the node set $V$ = $\mc{X} \cup \mc{D}$.
In ${G}$, each expert is not connected to any other experts, and the same is with the documents. A directed edge points from a document $d \in \mc{D}$ to an expert $x \in \mc{X}$, if $x$ has the authorship on $d$. 
This edge is denoted as ${e_{dx}}$. 
Thus, the set of edges $E$ contain directed edges from $\mc{D}$ to $\mc{X}$. 
Given ${e_{dx}}$, its weight, denoted as $w_{dx}$, comes from $\mb{DX}_{i(d),i(x)}$ (see Section~\ref{sec:notation}). 

\begin{figure}[!h]
    \begin{minipage}[t]{0.45\linewidth}
        \centering
        \includegraphics[trim=0cm 0cm 0cm 0cm, width=100pt]{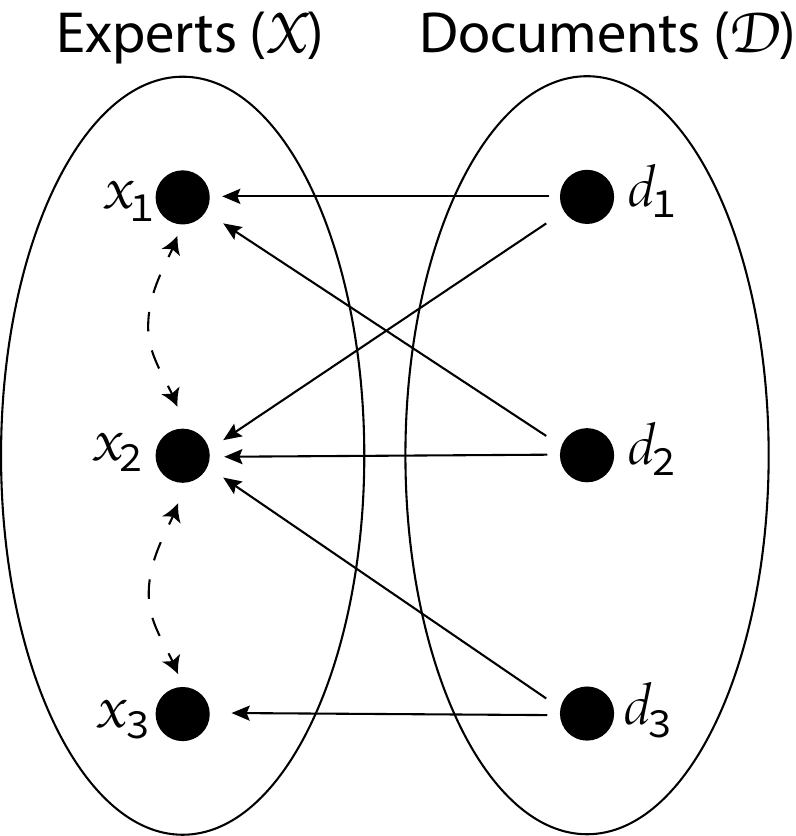}
        \subcaption{}
    \end{minipage}
    \quad
    \begin{minipage}[t]{0.45\linewidth}
        \centering
        \includegraphics[trim=0cm 0cm 0cm 0cm, clip, width=100pt]{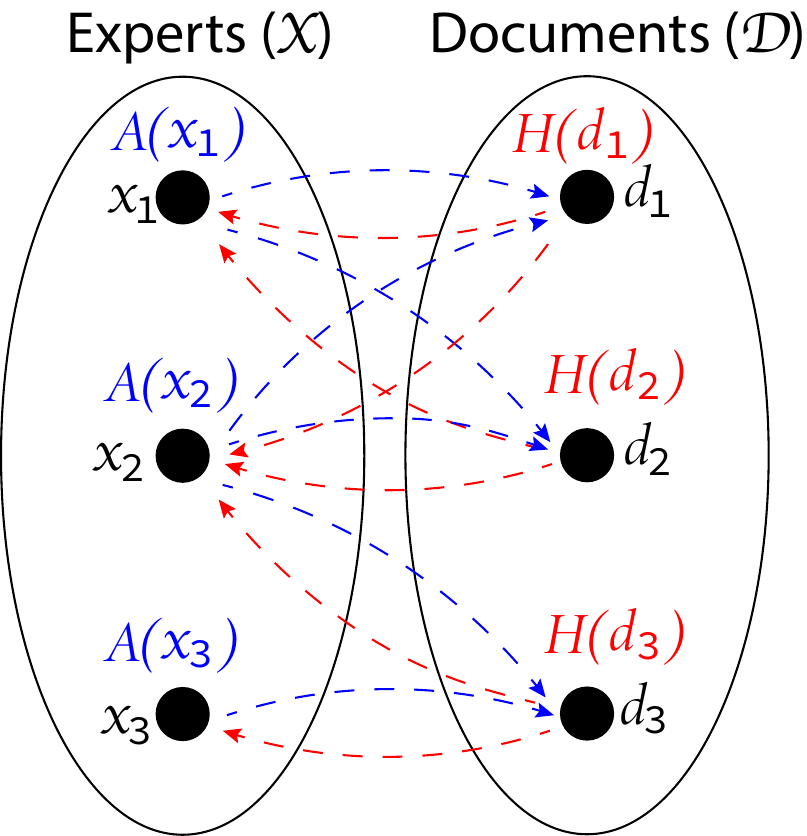}
        \subcaption{}
    \end{minipage}  
    \caption{An example \ecg (a) and the score propagation of authority and hub nodes (b).}\label{fig:ECG}
\end{figure}

An example \ecg is depicted in Fig.~\ref{fig:ECG}(a), where the solid lines show associations between experts and documents. The dashed arrows show implicit collaborations between experts via their joint documents: e.g.,
$x_1$ and $x_2$ have the joint documents $d_1$ and $d_2$, such that a collaboration between them is established as a bidirectional dashed arrow. 



\subsection{Reinforcing Expert Weights using \hits} \label{sec:cohits}

Note that \hits is a variation of CO-HITS~\citep{Hongbo:2009}. Thus, we first present the basic notion of CO-HITS on the structure of \ecg. That is, an important document is expected to point to important experts, while an important expert is linked by important documents. The importance of an expert $x$ is called the \textit{authority score} of $x$, and the importance of a document $d$ is called the \textit{hub score} of $d$. These scores are non-negative weights. Here, our goal is to reinforce the topic-sensitive weights of experts, estimated by $n$VSM, using \hits on the underlying \ecg.
For this, our idea is that given a topic $t$, we propagate the authority and hub scores with respect to $t$ by traversing $\mc{X}$ and $\mc{D}$ on the \ecg via an iterative process. 

An example is shown in Fig~\ref{fig:ECG}(b), 
where the hub scores, $H(d_1)$ and $H(d_2)$, are propagated to the expert $x_1$ to update the authority score $A(x_1)$; $H(d_2)$ is also propagate to update $A(x_2)$; and $H(d_3)$ is propagated to update $A(x_2)$ and $A(x_3)$. Once all the authority scores are updated, these scores are again propagated to the hubs to update their  scores. This performs iteratively. The intuition behind the iteration is the repeated mutual reinforcement to estimate authority and hub scores from co-linked nodes on the \ecg. 

In order for \EF to incorporate an topic into \hits, we take two steps. First, we extend the CO-HITS equation~\citep{Hongbo:2009} to accommodate a {topic}. We call this extension  \textit{topic-sensitive} CO-HITS. As the {initial authority and hub scores}, our {key idea } is to use the estimated topic-sensitive weights of experts and documents in $n$VSM, respectively. Second, we newly design and apply our variation of topic-sensitive CO-HITS into the \ecg. 
We elaborate these two steps in the rest of this section.

As the first step, we formally present the topic-sensitive CO-HITS equation, given an expert $x$ and a topic $t$:

\begin{align}\label{eq:hits_topic}
    \begin{split}
    A(x;t)^k &= (1-\lambda_{x}) \alpha_{x;t} + \lambda_{x}\sum_{e_{dx} \in E} w_{dx} H(d;t)^{k-1} \\
    H(d;t)^k &= (1-\lambda_{d}) \alpha_{d;t} + \lambda_{d}\sum_{e_{du} \in E} w_{du} A(u;t)^{k}
    \end{split}
\end{align}
where 
\begin{itemize}
    \item $A(x;t)^k$ and $H(d;t)^k$ are the topic-sensitive authority score of $x$ and topic-sensitive hub score of $d$, respectively, given $t$ at $k$-th iteration. 
    \item $w_{dx}$ denotes the weight of the edge $e_{dx}$, and thus  $w_{dx}=\mb{DX}_{i(d),i(x)}$ and $w_{du} = \mb{DX}_{i(d),i(u)}$, where $i(d)$ and $i(x)$ indicate the row and column index of $d$ and $x$, respectively, on $\mb{DX}$.
    \item $k$ indicates a iteration number staring from 1.
    \item $\alpha_{x;t}$ is the initial score for $A(x;t)^*$ and $\alpha_{d;t}$ is the initial score for $H(d;t)^*$ given $t$. 
    We call these scores \textit{personalised weights}.
    In this work, these personalised weights are normalised to be the widely used L2-norm~\cite{Kleinberg:1999}, that is, $\left(\sum_{x_i\in \mc{X}}
    \alpha_{x_i;t}\right)^{1/2}$ = $1$ and $\left(\sum_{d_i\in \mc{D}}
    \alpha_{d_i;t}\right)^{1/2}$ = $1$. Also, after updating the $k$-th iteration, the square root of the sum of squares of $A(x;t)^k$ and $H(d;t)^k$ are normalised using L2-norm, respectively. 
    Assigning the personalised weights provides crucial information in CO-HITS as they provide valuable and make an impact on the propagation of the updates of both authority and hub scores~\cite{Hongbo:2009}. 
    Our approach to determining the personalised weights is presented when discussing our proposed variation equation of Eq.~\ref{eq:hits_topic2}.
    
    
    \item $\lambda_{x}\in [0,1]$ and $\lambda_{d} \in [0,1]$  are  \textit{personalised parameter} for expert and document, respectively. These parameters determine how much we consider the personalised weights when calculating the $k$-th scores. Assigning lower values  indicates that higher importance is given to the personalised weights while reducing the propagation effects of co-linked nodes.
    
    \item Using Eq.~\ref{eq:hits_topic}, the topic-sensitive CO-HITS algorithm performs as follows: (1) with the personalised weights, a user-specified $k$ and a topic $t$, update all authority scores of $\mc{X}$; and (2) update all hub scores of $\mc{D}$. These steps are repeatedly performed $k$ times.
\end{itemize}



In the second step, we design the \hits equation and apply it on the underlying \ecg using two variation schemes of topic-sensitive CO-HITS:

\begin{equation}\label{eq:hits_topic2}
\scalemath{0.92}{
    \begin{split}
        A(x;t)^k &= (1-\lambda_{x}) A(x;t)^{k-1} + \lambda_{x}\left(\frac{\sum\limits_{e_{dx} \in E} w_{dx} H(d;t)^{k-1}}{\sum\limits_{e_{dx} \in E} w_{dx}} \right) \\
        H(d;t)^k &= (1-\lambda_{d}) H(d;t)^{k-1} + \lambda_{d} \left(\frac{\sum\limits_{e_{du} \in E} w_{du} A(u;t)^{k}}{\sum\limits_{e_{du} \in E} w_{du}} \right)
    \end{split}
}
\end{equation}

where the interpretation of all the variables is the same as presented for Eq.~\ref{eq:hits_topic}, except the following two variation schemes. 

The first variation scheme is that rather than using the \textit{fixed} personalised weights $\alpha_{x;t}$ and $\alpha_{d;t}$, \hits uses \textit{dynamic} personalised weights $A(x;t)^{k-1}$ and $H(d;t)^{k-1}$ at each $k$-th iteration. In Eq.~\ref{eq:hits_topic}, regardless of iterations, the authority and hub scores at each iteration are fixed to be $\alpha_{x;t}$ and $\alpha_{d;t}$. Different from it, our approach is to use personalised weights at the $k$-th iteration as the ($k$-1)-th authority and hub scores. By doing so, in the calculation of the authority (resp. hub)  scores at the $k$-th iteration, the our aim is to exploit 
both the propagation of the hub (resp. authority) scores and the effect of the authority (resp. hub) score at the ($k$-1)-th iteration. 
Thus, in \hits, personalised weights are updated at each iteration based on the authority and hub scores at the previous iteration. In our approach, as the initial personalised weights, we use the topic-sensitive weights of experts and documents estimated using $n$TFIDF in $n$VSM. Thus, $A(x;t)^{0}$ = $\mb{TX}_{i(t),i(x)}$ and $H(d;t)^{0}$ = $\mb{TD}_{i(t),i(d)}$.
Similarly, in the topic-sensitive CO-HITS equation in Eq.~\ref{eq:hits_topic}, $\alpha_{x;t}$ and $\alpha_{d;t}$ are set to be $A(x;t)^{0}$ and $H(d;t)^{0}$, respectively.
By doing so, we integrate $n$VSM  with \hits, generating a new unified formula for this integration. Our intuition for this integration is to improve
the accuracy for expert finding by further exploring the implicit relationships between experts, derived from the \ecg, in addition to the results of the $n$VSM approach. Note that $n$VSM  ignores such relationships, only utilising the importance of a document $d$; the importance of a topic $t$ from the documents of an expert $x$; and the importance of $x$ given $d$ (see Eq.~\ref{eq:pw_2}).
    
The second variation scheme is that the \textit{aggregation} of the authority and hub scores is different from that of topic-sensitive CO-HITS. In Eq.~\ref{eq:hits_topic}, $A(x;t)^{k}$ and $H(d;t)^{k}$  are calculated based on the square root of the \textit{sum} of squares of $H(d;t)^{k-1}$ and $A(x;t)^{k}$, respectively. This approach tends to assign a higher authority score to an expert $x$ who has \textit{more} documents (i.e. $|\mc{D}_x|$). Similarly, it is likely that a higher hub score is given  to a document $d$ that is linked to \textit{more} experts (i.e. $|\mc{X}_d|$) that have $d$.

Instead, in \hits, we use the \textit{central tendency} of $H(d;t)^{k-1}$ to calculate $A(x;t)^{k}$; and also use the \textit{central tendency} of $A(x;t)^{k}$ to calculate $H(d;t)^{k}$. The `average' is used to measure such central tendency. The  reason is that we have already incorporated the idea of using `the sum of squares of authority and hub scores', used in topic-sensitive CO-HITS, in the context of $n$VSM. Note that in $n$VSM, we calculated the topic-sensitive weights of experts by using the \textit{sum} operator as seen in Eq.~\ref{eq:pw_2} (i.e. $\sum_{d \in \mc{D}_{x}}$). Thus, to avoid the duplicated use of this `sum' operator, given a topic $t$, we design \hits in a way that estimates the importance of an expert $x$ at the $k$-th iteration (i.e. $A(x;t)^k$) by calculating the \textit{average} of the ($k$-1)-th hub scores, in addition to personalised weight $A(x;t)^{k-1}$. Similarly, we estimate the importance of a document $d$ (i.e. $H(d;t)^k$) by calculating the \textit{average} of the $k$-th authority scores, in addition to personalised weight $H(d;t)^{k-1}$. In the name \hits, `$\mu$' indicates the `average' so that \hits means a particular topic-specific CO-HITS using the 'average' importance of authority and hub scores.

We also highlight other features of \hits. First, as with  topic-sensitive CO-HITS, the updated authority and hub scores at each iteration are normalised using L2-norm.
Second, if $\lambda_x$ and $\lambda_d$ are 0, \hits returns the initial personalised weights at each iteration. Thus, \EF does not use the score propagation effects on the \ecg, returning the same results obtained from $n$VSM. Third, if $\lambda_x$ and $\lambda_d$ are all equal to 1, \hits does not incorporate personalised weights. However, it calculates $H(x;t)^{1}$ based on the $H(d;t)^{0}$ that was obtained from the topic-document matrix $\mb{TD}$, i.e.,  $H(d;t)^{0}$ = $\mb{TD}_{i(t),i(d)}$, generated by $n$VSM. Also, $H(d;t)^{1}$ is calculated based on $A(u;t)^{1}$.

\section{Evaluation of \EF}\label{sec:eval}
To assess the effectiveness of \EF, we conduct the following evaluation. First, we measure the effectiveness of the first knowledge facet of \EF: $n$VSM (Section~\ref{sec:weight_eval}).
Second, we show how to empirically find the best values for personalised parameters of \EF (Section~\ref{sec:lamb_eval}). 
Third, we evaluate that \EF is a highly competitive model for expert finding, in comparison with $n$VSM and the two GM approaches~\citep{Gollapalli:2013,Daniel:2015}. Further, to show the capability of the second knowledge facet of \EF, i.e., \hits, over topic-sensitive CO-HITS (simply CO-HITS), we compare \EF with an alternative \EF form that combines $n$VSM and CO-HITS (Section~\ref{sec:eval_ef}). Finally, we summarise our evaluation (Section~\ref{sec:discussion}).

\subsection{Datasets}\label{sec:data}

We use four benchmark datasets in our evaluation. One is the {Lattes Expertise Retrieval} ({{\textit{LExR}}})\footnote{LExR is available to download from \url{http://toinebogers.com/?page_id=240}} test collection~\citep{Vitor:2016} for expertise retrieval in academic. LExR provides a comprehensive, large-scale benchmark for evaluating expertise retrieval and it covers all knowledge areas (e.g. earth sciences, biology, health sciences, languages, art, etc) working in research institutions all over Brazil. Most publications are written in Portuguese, Spanish and English. In our evaluation, we only consider the English documents for our readability. 
The other three datasets\footnote{These datasets can be downloaded from  \url{http://www.lbd.dcc.ufmg.br/lbd/collections}.} are Information Retrieval (IR), Semantic Web (SW), and Computational Linguistics (CL) which are filtered subsets of DBLP dataset \citep{bordea2013benchmarking}. 
In these four datasets, we regard the authors as experts $\mc{X}$ and the publications as documents $\mc{D}$, where each publication is seen as a mixture of title and abstract. From $\mc{D}$, we extract phrases as the first step in \EF (Section~\ref{sec:overview}). 

These datasets also provide the ground-truth about who are the known experts for the  known topics. The expert degrees for each topic are represented as non-relevance, relevance, and high relevance in LExR. We regard individuals with non-relevance as \textit{non-experts}, and individuals with relevance and high relevance as \textit{experts}.
IR, SW and CL also provide the expert list for each topic. We formalise the candidates in such list as \textit{experts}, and otherwise \textit{non-experts}.

From each dataset, we preprocess the following steps to be used in our evaluation. First, we remove publications containing empty title and abstract. Second, we remove publications whose abstracts provide little information, that is, less than 5 words after removing stopwords. 
Third, if there exists duplicated topics, we remove such ones.
Table \ref{tab:data_desc} shows an overview of the datasets after performing these steps.

\captionsetup[table]{labelfont=bf, skip=0pt}
\begin{table}[!h]
    \caption{A summary of our four datasets}
    \begin{center}
        \scalebox{0.87}{
            \setlength{\tabcolsep}{4pt}
            \begin{tabular}{p{5cm}rrrr}
            \hline
            & LExR & IR & SW & CL \\
            \hline
            \# of documents & 14879 & 2355 & 1519 & 1667 \\
            \# of experts & 620 & 276 & 394 & 358 \\
            \# of topics & 227 & 268 & 2046 & 1583 \\
            Avg. \# of documents per expert & 28 & 9 & 4 & 5 \\
            Avg. \# of experts per topic & 6 & 10 & 9 & 8 \\
            Median \# of experts per topic & 5  & 8 & 6 & 5 \\
            Max \# of experts per topic & 26  & 177 & 226 & 158 \\
            \hline
            \end{tabular}
        }
    \end{center}
    \label{tab:data_desc}
\end{table}

We note that our chosen datasets are relatively more comprehensive than some previous works, which focused on academic domains for their evaluation, in terms of the number of topics considered, thereby providing a reasonable measure of the effectiveness of \EF. For example, the works \cite{Deng:2008}, \cite{Gollapalli:2013} and \cite{Wang:2015} used two datasets with seven topics, two datasets with 13 and 203 topics and one dataset with 14 topics, respectively\footnote{These datasets are also no longer publicly available.}. Note that our evaluation
have been done using the larger numbers of the topics on the four datasets as seen in Table~\ref{tab:data_desc}.

\begin{table*}[!t]
    \caption{Expertise topics and corresponding similar phrases in four datasets}
    \begin{center}\label{tab:sim_pt}
        \scalebox{0.6}{
            \begin{tabular}{cc|cc|cc|cc}
            \toprule
            \multicolumn{2}{c}{LExR} & \multicolumn{2}{c}{IR}  & \multicolumn{2}{c}{SW}  & \multicolumn{2}{c}{CL} \\
            \cmidrule{1-8}
            Topic & Phrase & Topic & Phrase & Topic & Phrase & Topic & Phrase \\
            \midrule
            synthesis & synthesis & information retrieval & information retrieval & semantic web & semantic web & question answering & question answering  \\
            risk factor & risk factor &  search engine & search engine & linked data & linked data & knowledge transfer & knowledge transfer \\
            public health & public health & patent search & patent search & text understanding & text understanding & summarization & summarization \\
            thin film & ultrathin film & data modelling & information modelling & auction & bidding & machine vision & computer vision \\
            development validation & validation process & cooperative work & collaborative working & invalidity & inadequacy & image classification & image recognition \\
            \bottomrule
            \end{tabular}            
        }
    \end{center}
\vspace{-10pt}
\end{table*}


\subsection{Evaluation Framework}\label{sec:eval_frame}
We present our evaluation configuration and metrics. Recall that as a topic, we use a phrase. We assume that the maximum word length of each phrase is 3 in our evaluation. Also, we observe that there is no guarantee that an original known topic $t_g$ always appears in documents $\mc{D}$ in each dataset. 
Thus, given each $t_g$, we find its most similar phrase $t$ from $\mc{D}$. Then, $t$ is alternatively used as a topic, instead of $t_g$.
To find $t$ given $t_g$, we use the scientific pre-trained model SciBERT\footnote{The pre-trained SciBERT model can be downloaded at \url{https://github.com/allenai/scibert}}~\citep{Beltagy2019SciBERT} that is a scientific language model trained on the fulltext of 1.14M papers and 3.1B words, where the papers were collected from `semanticscholar.org'.
Using this model allows us to measure a semantic similarity between $t_g$ and $t$ by their cosine similarity according to their corresponding vectors represented in the model. More specifically, assume that $s_1$ is an original known topic and $s_2$ is a phrase extracted from $\mc{D}$. Then, we measure their similarity as:

\begin{equation}
    sim(s_1, s_2) \approx \cos(\vec{s_{1}}, \vec{s_{2}}) = \frac{\vec{s_{1}}^{\,} \cdot \vec{s_{2}}^{\,} }{\left\Vert \vec{s_{1}}^{\,} \right\Vert \left\Vert \vec{s_{2}}^{\,} \right\Vert},
\end{equation}
where $\vec{s_1}$ and $\vec{s_2}$ are the represented vectors of $s_1$ and $s_2$ in SciBERT, respectively. Each of these vectors is estimated by the average of the embedded vectors of its constituent terms. 
Suppose that $s_1$ consists of $n$-terms, $s_1 = (w_1, \ldots, w_n)$, then,
$\vec{s_1} \approx \frac{1}{n} (\vec{w_1} + \ldots + \vec{w_n}),$
where $(\vec{w}_1, \ldots, \vec{w}_n)$ are the embedded vectors of $(w_1, \ldots, w_n)$. The same principle is applied to $s_2$. Table~\ref{tab:sim_pt} shows the  examples of five topic-phrase pairs  in each dataset, where each pair shows an original known topic $t_g$ and the most similar phrase $t$ used as a topic in our evaluation. As we see, some phrases ($t$) are equal to the original topic ($t_g$), while some others phrases are semantically very similar to the corresponding original topic (e.g. `image classification'-`image recognition' on CL).

Other evaluation configuration includes: (1) we assume that the importance of documents is the same (i.e. $p(d)$=1) and the importance of all experts of $d$ is the same (i.e. $p(x|d)$=1) in Eq.~\ref{eq:pw_2}. The reason is that one of our primary focuses is to evaluate the capability of $n$TFIDF in $n$VSM in calculating $p(t|d)$ in Eq.~\ref{eq:pw_2};
(2) Thus, we also fix $w_{dx}$=1 and $w_{du}$=1 in Eq.~\ref{eq:hits_topic} and Eq.~\ref{eq:hits_topic2}; and (3) from our empirical testing, we observed Eq. \ref{eq:hits_topic} and Eq. \ref{eq:hits_topic2} are commonly converged after 5 iterations, so we set $k = 5$.

For all expert finding models in our evaluation, our aim is to generate a ranked list of experts based on the outcome of the formula: (1) Eq.~\ref{eq:hits_topic2} in \EF, (2) Eq.~\ref{eq:hits_topic} in CO-HITS, (3) Eq.~\ref{eq:pw_2} and Eq.~\ref{eq:topic_by_terms} in DLM, (4) Eq.~\ref{eq:pw_2} and Eq.~\ref{eq:entropy2} in TFIDF-based VSM,  and (5) Eq.~\ref{eq:pw_2} and Eq.~\ref{eq:ntfidf} in $n$VSM, where all these methods are compared, in addition to WISER~\cite{WISER:2019} and RepModel~\cite{Daniel:2015}.

We use two widely-used evaluation metrics for expert finding~\citep{Wang:2015, Hongbo:2009}:
(1) {\it precision at rank n (P@n)} and (2) {\it Mean Average Precision (MAP)}. P@n measures the relevance of the $n$-top ranked experts with respect to a given query topic, defined as~\citep{Hongbo:2009}:
\begin{equation}
    P@n = {
    |S_{n} \cap R_{t}|
    }/{n},
\end{equation}
where $S_{n}$ is the set of $n$-top recommended experts for a given topic $t$, and $R_{t}$ is the set of known experts for $t$. We report from P@10 to P@30 (increasing by 5) for each topic and take the average over all topics. MAP measures the overall ability of a method to differentiate between known experts and non-experts. The \textit{average precision (AP)} is defined as~\citep{Wang:2015}:
\begin{equation}\label{eq:AP}
    AP = \frac{
    \sum_{i=1}^{n}(P@i * rel(i))
    }{|R_{t}|}
\end{equation}
where $i$ is the rank, $rel(i)$ is a binary function indicating 1, if the result at $i$ is a known expert, otherwise 0. MAP is the mean value of AP values over all topics, and we use $n = 30$ as used in~\cite{Deng:2008}.

\begin{figure*}[!t]
\begin{center}
    \begin{tabular}{llll}
      \includegraphics[trim=1.8cm 0cm 0cm 0cm,width=39mm,height=40mm]{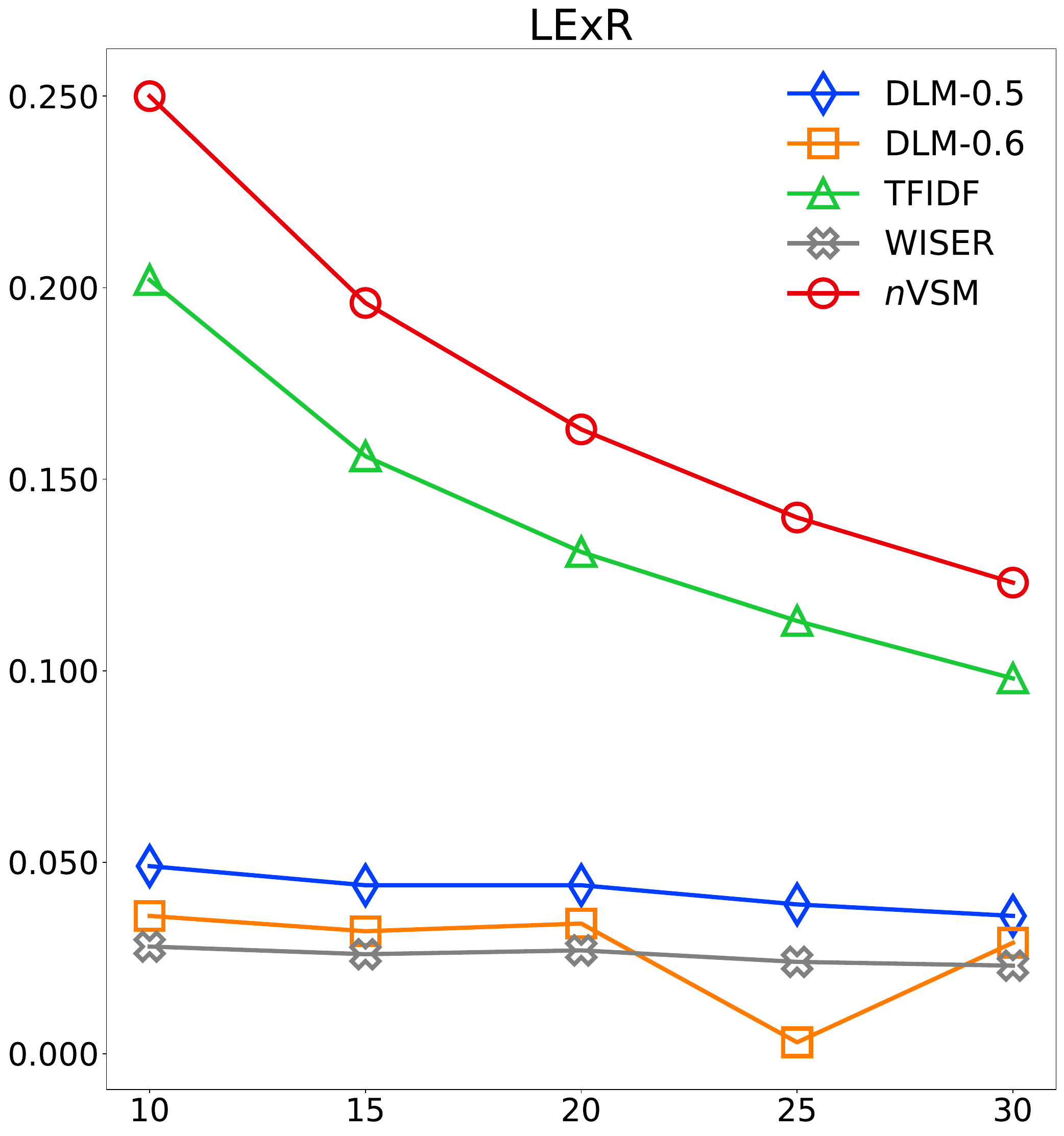} &  \includegraphics[trim=1.8cm 0cm 0cm 0cm,width=39mm,height=40mm]{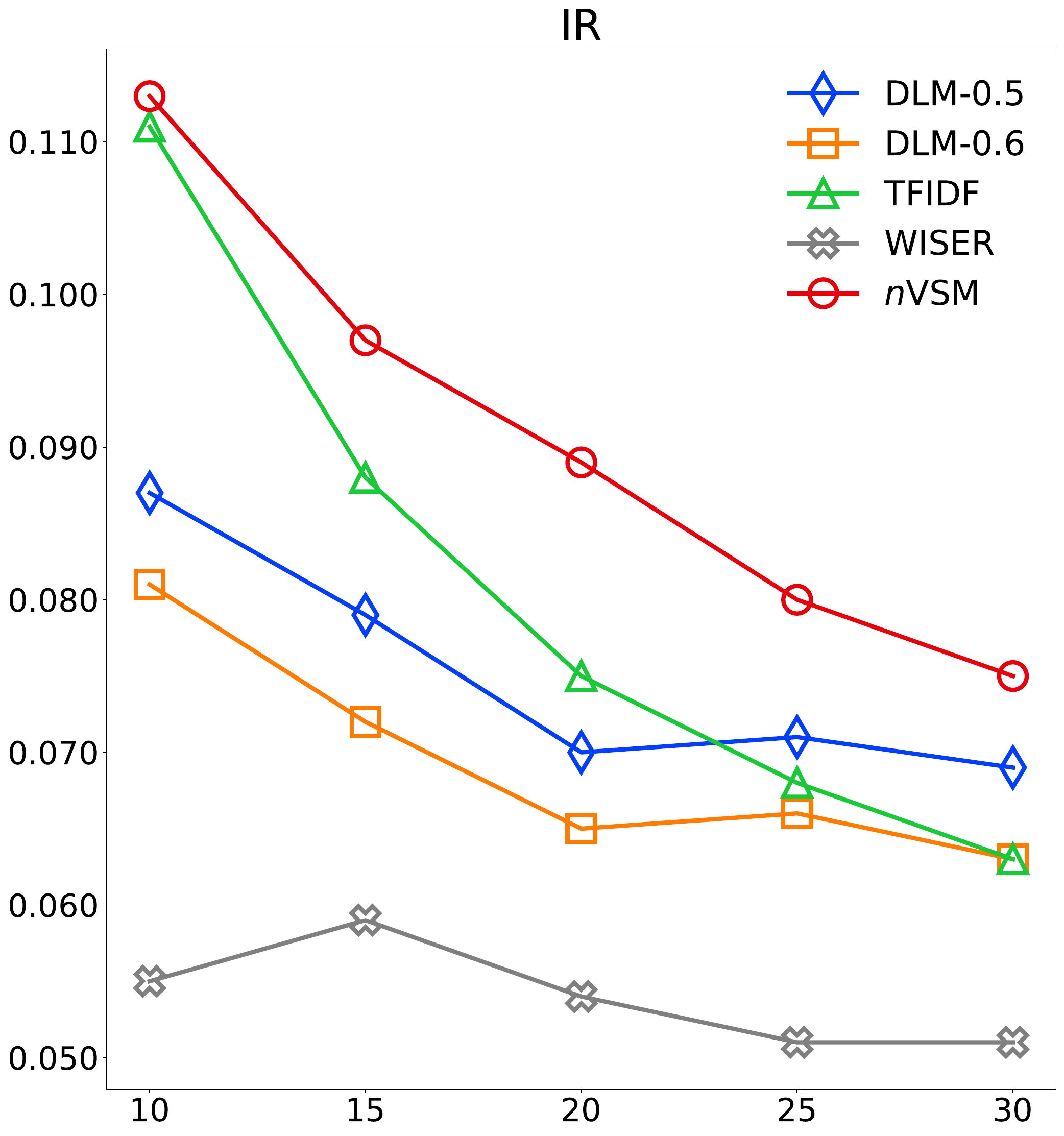} &  \includegraphics[trim=1.8cm 0cm 0cm 0cm,width=39mm,height=40mm]{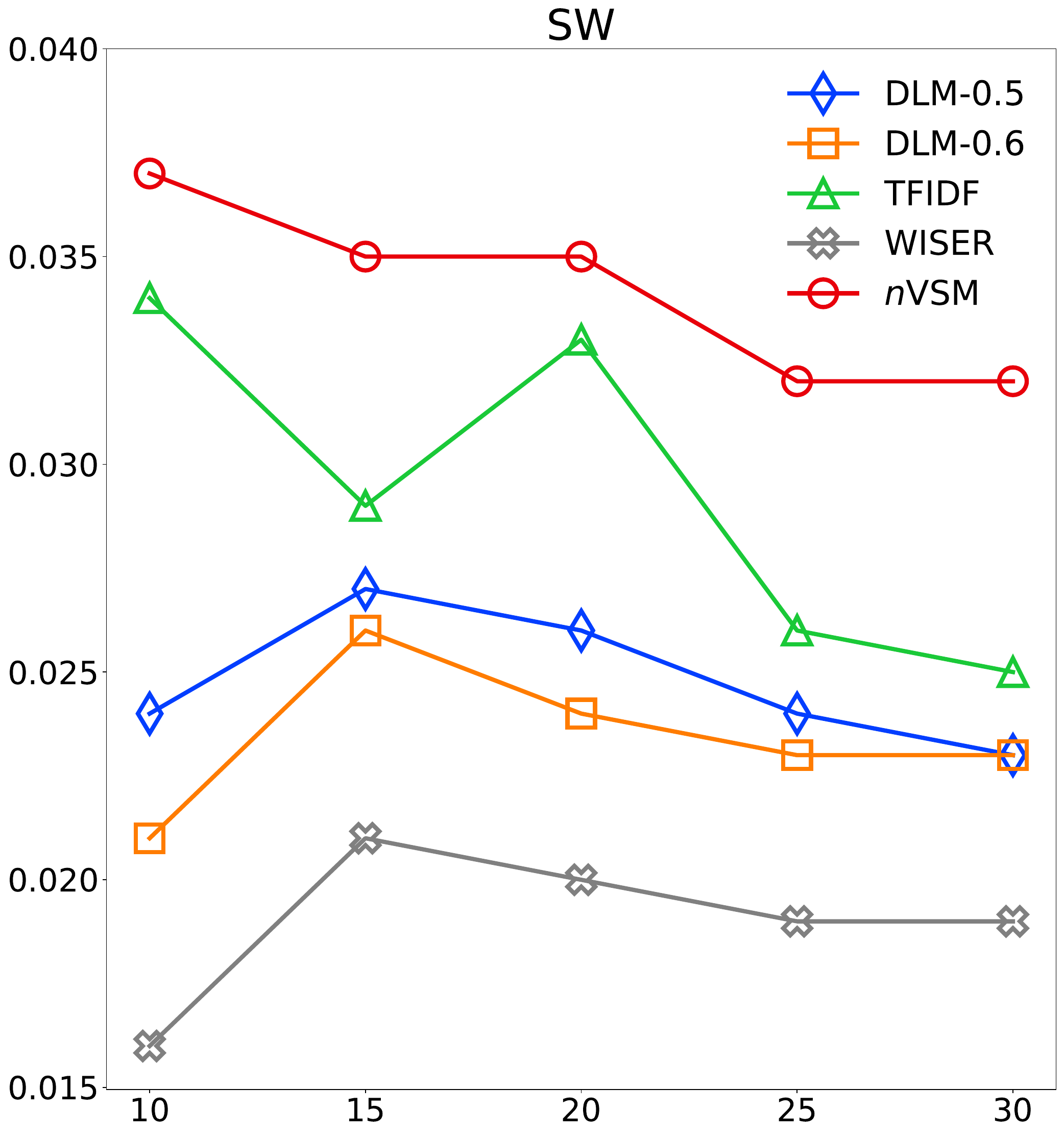} &  \includegraphics[trim=1.8cm 0cm 0cm 0cm,width=39mm,height=40mm]{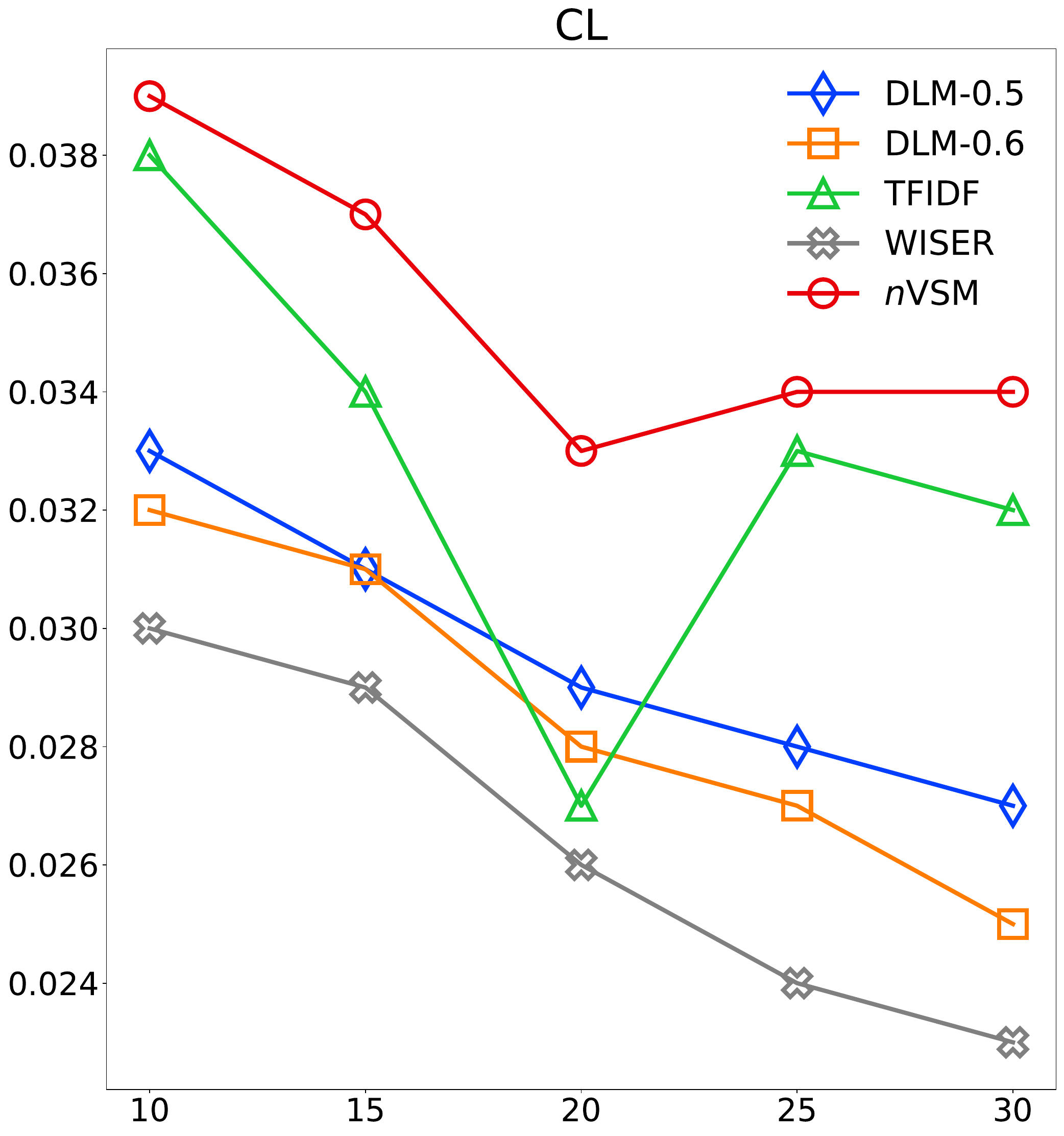}
    \end{tabular}
    \caption{Comparison on the average precision (AP) values: x-axis shows n of P@n and y-axis shows AP values.}
    \label{tab:eval_APC}
\end{center}
\end{figure*}

\captionsetup[table]{labelfont=bf, skip=0pt}
\begin{table*}[!t]
    \caption{MAP and the the improvement ratio of $n$VSM.}
    \begin{center}
        \scalebox{0.9}{
            \setlength{\tabcolsep}{9pt}
            \begin{tabular}{l r l r l r l r l r l}
            \hline
            & LExR & & IR & &  SW & & CL & & Avg. \\
            \hline
            DLM-0.5 & 0.200 & (233.0\%) & 0.208 & (6.7\%) & 0.070 & (51.4\%) & 0.063 & (68.3\%) & 0.135 & (103.7\%) \\
            DLM-0.6 & 0.159 & (318.9\%) & 0.185 & (7.8\%) & 0.068 & (55.9\%) & 0.058 & (82.8\%)  & 0.118 & (133.1\%) \\
            TFIDF & 0.493 & (35.1\%) & 0.206 & (7.8\%) & 0.087 & (21.8\%) & \textbf{0.120} & (-11.7\%) & 0.226 & (21.7\%) \\
            WISER & 0.117 & (469.2\%) & 0.150 & (48.0\%) & 0.057 & (86.0\%) & 0.051 & (107.8\%) & 0.094 & (192.6\%) \\
            $n$VSM & \textbf{0.666} &  & \textbf{0.222} & & \textbf{0.106} &  & 0.106 &  & \textbf{0.275} & \\
            \hline
            \end{tabular}
        }
    \end{center}
    \label{tab:eval_MAPC}
\vspace{-10pt}
\end{table*}

\subsection{Evaluation of $n$VSM} \label{sec:weight_eval}
As $n$VSM is one key compoenent in \EF, we first measure its effectiveness. As presented in Section~\ref{sec:ntfidf}, the concepts VSM and DLM are closely related. Thus, we compare $n$VSM with TFIDF-based VSM and two particular DLMs:
(1)
\textit{{TFIDF}-based VSM} expressed using Eq.~\ref{eq:pw_2} and Eq. \ref{eq:entropy2} (denoted as TFIDF); (2) \textit{The {DLM} model}~\citep{Balog:2009,Wang:2015} denoted using Eq.~\ref{eq:pw_2} and Eq.~\ref{eq:topic_by_terms} in which the probability of individual terms is estimated by Eq.~\ref{eq:document_model}, where we use two values for the best $\lambda_\theta$: 0.5 (\textit{DLM-0.5}) and 0.6 (\textit{DLM-0.6}), as suggested by \citep{Balog:2009} and \citep{Wang:2015}, respectively; and
(3) A recent probabilistic model {\textit{WISER}}~\cite{WISER:2019} that combines the document-centric approach exploiting the occurrence of topics in experts' documents, with the profile-centric approach computing relatedness between experts using an external knowledge source, Wikipedia. Since our work does not consider such an external knowledge source, we only consider WISER with the document-centric approach for the fair comparison. In WISER,
the topic-sensitive weight of an expert $x$ given a topic $t$ is calculated using Reciprocal Rank~\cite{Macdonald:2006}: $\sum_{d}^{\mc{D}_{x, t}}$ $\frac{1}{rank(d)}$ that represents the ranks of $x$'s documents where $t$ appears ($\mc{D}_{x, t}$). Since $t$ is a phrase, $\mc{D}_{x, t}$ consists of the subset of $\mc{D}_x$ that any of $t$'s constituent terms appears. $rank(d)$ is the ranking position of a document $d$ out of $\mc{D}$, where the position is determined by BM25~\citep{BM25:2009}. The hyper-parameters $k1$ and $b$ in BM25 are set to be 1.2 and 0.75, respectively, based on the suggestion~\cite{Yuanhua:2011}. 

The evaluation results are presented in Fig. \ref{tab:eval_APC} that
shows the AP values with n (n=10, 15, $\ldots$, 30) of P@n for all topics. We observe the following: (1) Overall, the VSM approaches (TFIDF and $n$VSM) largely outperform all DLM-0.5, DLM-0.6 and WISER. This  indicates the VSM approaches can be more effectively used for identifying topic-sensitive experts than the compared DLMs; (2) DLM-0.5 is consistently better than DLM-0.6 but their difference seems minor; and (3) $n$VSM is clearly better than TFIDF from P@10 to P@30 consistently over all the four datasets. Also, $n$VSM substantially outperforms all the compared methods on all the four datasets. 
Table \ref{tab:eval_MAPC} shows the results on MAP and the relative improvement ratio of $n$VSM over the other models. The best one in each dataset is denoted in boldface. We see that $n$VSM's improvements over DLM-0.5 and DLM-0.6 are substantial: up to 318.9\% over DLM-0.6 on LExR, 7.8\% over DLM-0.6 on IR, 55.9\% on CL, and 82.8\% on CL. Moreover, $n$VSM substantially outperforms WISER from 48.0\% on IR to 469.2\% on LExR. Also, $n$VSM is highly better than TFIDF except the only one case on CL.
On average, we observe that $n$VSM largely outperforms DLM-0.5 in 103.7\%; DLM-0.6 in 133.1\%; WISER in 192.6\%; and TFIDF in 21.7\% across the four datasets. 
In summary, the results show an empirical evidence that $n$VSM can be competitive and effectively used for expert finding. Also, these show that \EF is equipped with a powerful component, $n$VSM, for expert finding.


\subsection{Finding the Best Values for Personalised Parameters in \EF: $\lambda_x$ and $\lambda_d$} \label{sec:lamb_eval}
We now present how to empirically find the best values for personalised parameters  $\lambda_{x}$ and $\lambda_{d}$ of \hits (see Eq.~\ref{eq:hits_topic2}) which is another key component of \EF. Our approach is to make a full use of all the four datasets to determine such values. For this, we measure the \textit{mean} impact of different values of $\lambda_{x}$ and $\lambda_{d}$, respectively, on generating the MAP results from the four datasets. Our aim is to provide an empirical guideline for choosing the best values for these parameters. 
Formally, let $Z$ be the set of candidate values [0, 0.1, $\ldots$, 1.0] for $\lambda_{x}$ and $\lambda_{d}$. Then, let us define MAP$(a, b)$ as the MAP value using a pair of $a\in Z$ for $\lambda_{x}$ and $b\in Z$ for $\lambda_{d}$. 
First, we choose the best value for $\lambda_{x}$. To this end, for each value $a \in Z$, we compute the mean of the MAP values with all values in $Z$ in each dataset:

\begin{equation}\label{eq:avg_lambda_x}
Avg(a, \lambda_{x}) = \frac{1}{|Z|}\sum_{b \in Z}{\textrm {MAP}} (a, b).
\end{equation}
Then, we obtain the $|Z|$-length vector of $Avg(a, \lambda_{x})$ for all values in $Z$. Let us say that this vector is denoted as $Avg(Z, \lambda_x)$. For example, if $Avg(Z, \lambda_x) = [1, 0.9, 0.8, \ldots, 0]$, then the corresponding element-wise rank vector is $R(Avg(Z, \lambda_x)) = [11, 10, 9, \ldots, 1]$, where the higher rank indicates the higher mean of the MAP values. Similarly, we use $R^i(Avg(Z, \lambda_x))$ to denote the $R(Avg(Z, \lambda_x))$ calculated on the dataset $i$. Finally, we compute the element-wise mean rank across the four datasets:

\begin{equation}\label{eq:avg_r_lambda_x}
    AvgR(Z,\lambda_{x})) = \frac{1}{n}\sum_{i=1}^{n}{R^i(Avg(Z,\lambda_{x}))},
\end{equation}
where $n = 4$ corresponding to the number of datasets. Using the above equation, we find the best value for $\lambda_{x}$ that is the $a \in Z$ generating the highest rank.

\begin{figure}[!h]
    \begin{minipage}[b]{0.45\linewidth}
        \centering
        \includegraphics[trim=0cm 0cm 0cm 0cm, width=105pt]{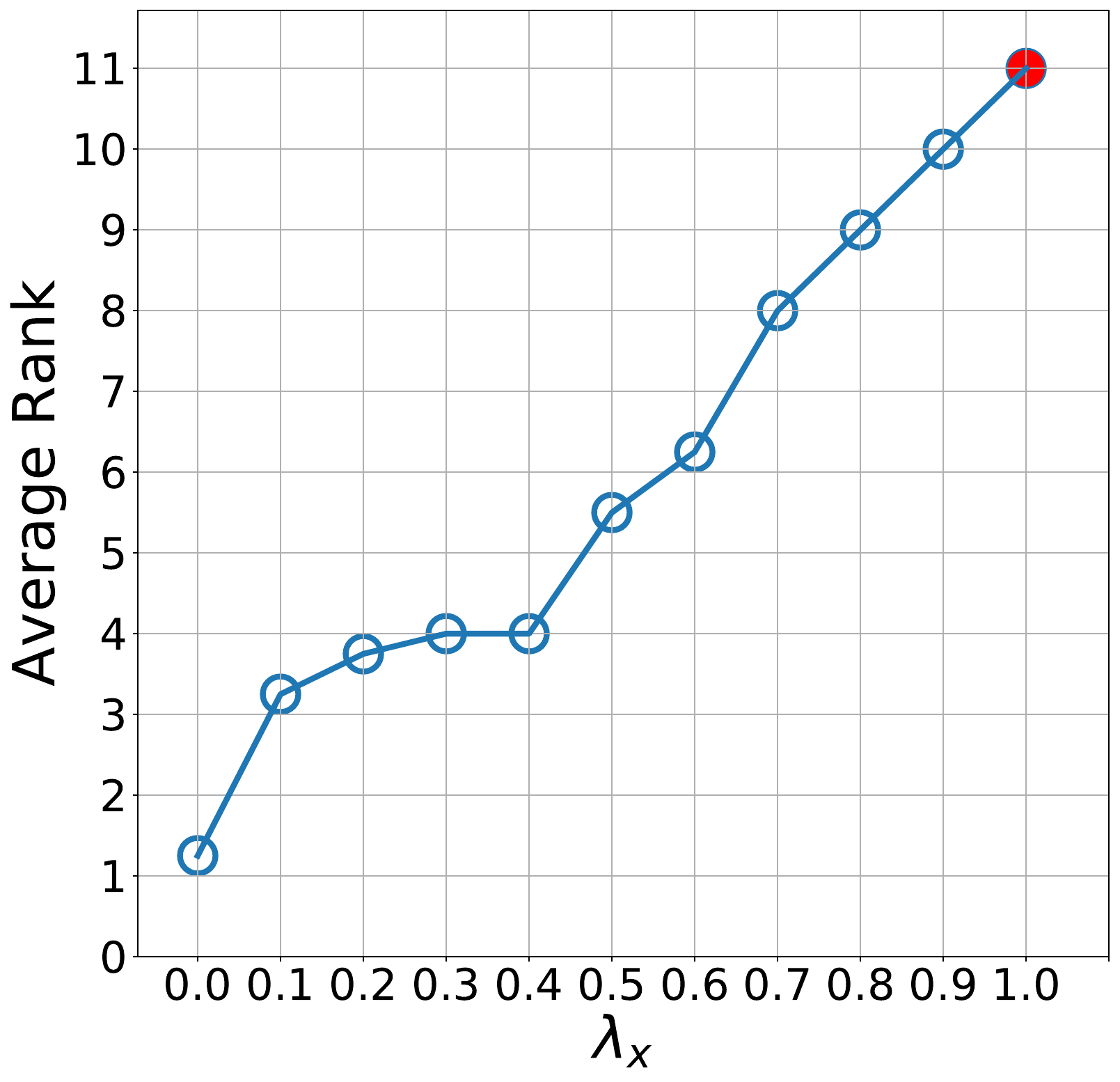}
        \subcaption{}
    \end{minipage}
    \quad
    \begin{minipage}[b]{0.45\linewidth}\vspace{0pt}
        \centering
        \includegraphics[trim=0cm 0cm 0cm 0cm, clip, width=105pt]{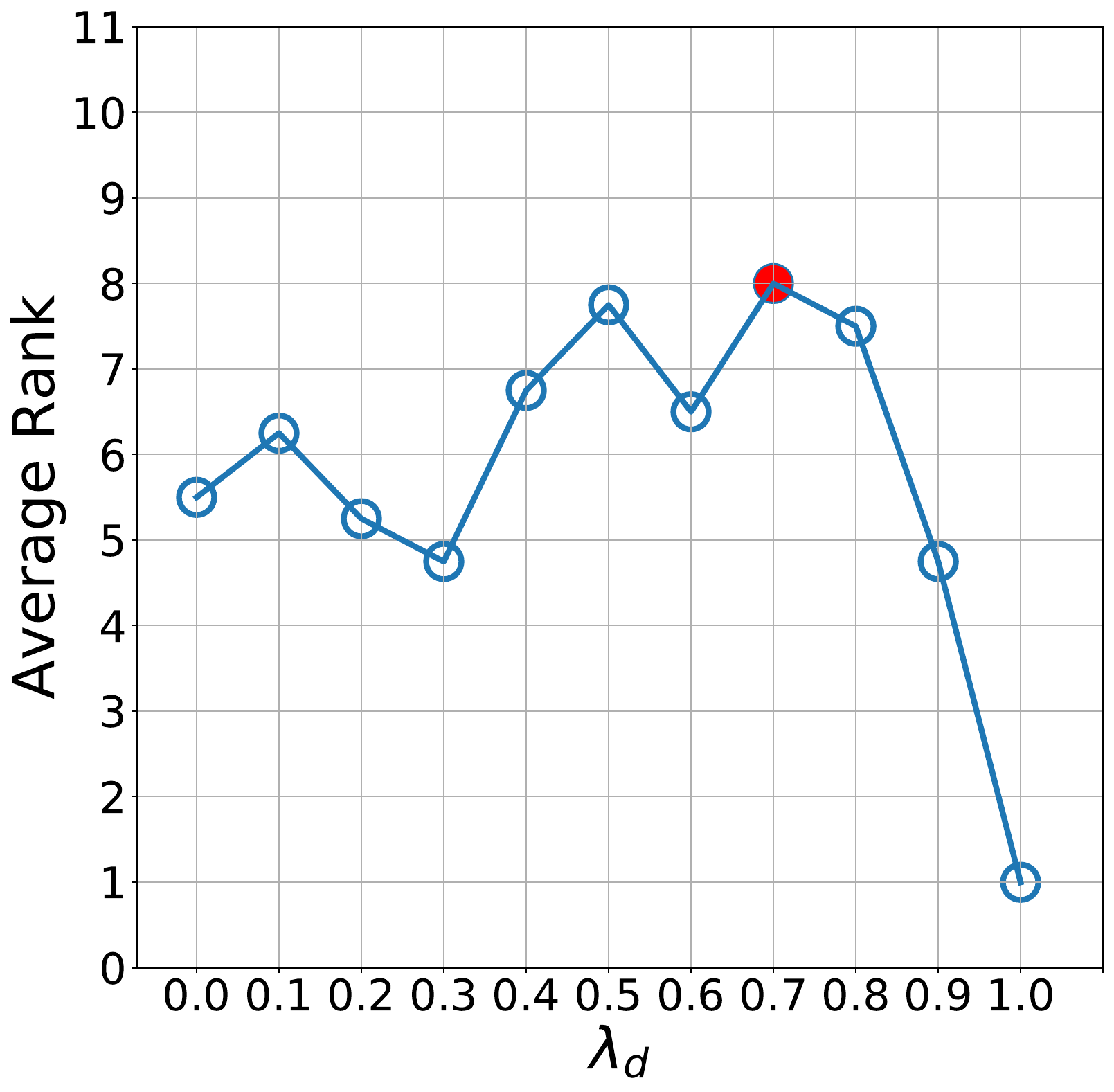}
        \subcaption{}
    \end{minipage}  
    \caption{Finding the best values for $\lambda_{x}$ and $\lambda_{d}$.}
    \label{fig:eval_lamb_XP}
\vspace{-10pt}
\end{figure}

Finding the best value for $\lambda_d$ is the same as the above procedure, except that we fix $a$ to be the identified best value for $\lambda_x$.
Thus, Eq.~\ref{eq:avg_lambda_x} is modified as:
$Avg(b, \lambda_{d}) = {\textrm {MAP}} (a, b)$.
Then, we obtain the $|Z|$-length vector of $Avg(b, \lambda_{d})$ for all possible values for $b \in Z$. This vector is denoted as $Avg(Z, \lambda_d)$. Also, $R^i(Avg(Z, \lambda_d))$ indicates the $R(Avg(Z, \lambda_d))$ calculated on the dataset $i$. Finally, we compute the element-wise mean rank across the four datasets using the Eq.~\ref{eq:avg_r_lambda_x} except that $\lambda_x$ is replaced with $\lambda_d$.
By doing so, we find the best value for $\lambda_{d}$ by choosing the $b \in Z$ generating the highest rank.
Fig.~\ref{fig:eval_lamb_XP}(a) - (b) show the average ranks of values in $Z$ for $\lambda_{x}$ and $\lambda_{d}$, respectively, across four datasets. As we see, $\lambda_{x}$ = $1.0$ produces the highest rank, whereas $\lambda_{d}$ = $0.7$ is the highest rank with $\lambda_{x}$ = $1.0$. The best ones are denoted in red color. 


\subsection{Evaluation of \EF}\label{sec:eval_ef}
We now evaluate \EF using the best values for $\lambda_{x}$ and $\lambda_{d}$. To measure its relative effectiveness, we also compare it with $n$VSM as well as two GMs: \textit{ADT}~\citep{Gollapalli:2013} and \textit{Reputation Model} (simply \textit{RepModel})~\cite{Daniel:2015}. Further, we compare \hits (Eq.~\ref{eq:hits_topic2}) with CO-HITS (Eq.~\ref{eq:hits_topic}) to validate the stronger capability of \hits's variation approach over CO-HITS. Finally, we show that \EF works well regardless of \textit{topic coverage}.

\begin{figure*}[!t]
\begin{center}
    \begin{tabular}{llll}
      \includegraphics[trim=1.8cm 0cm 0cm 0cm,width=39mm,height=40mm]{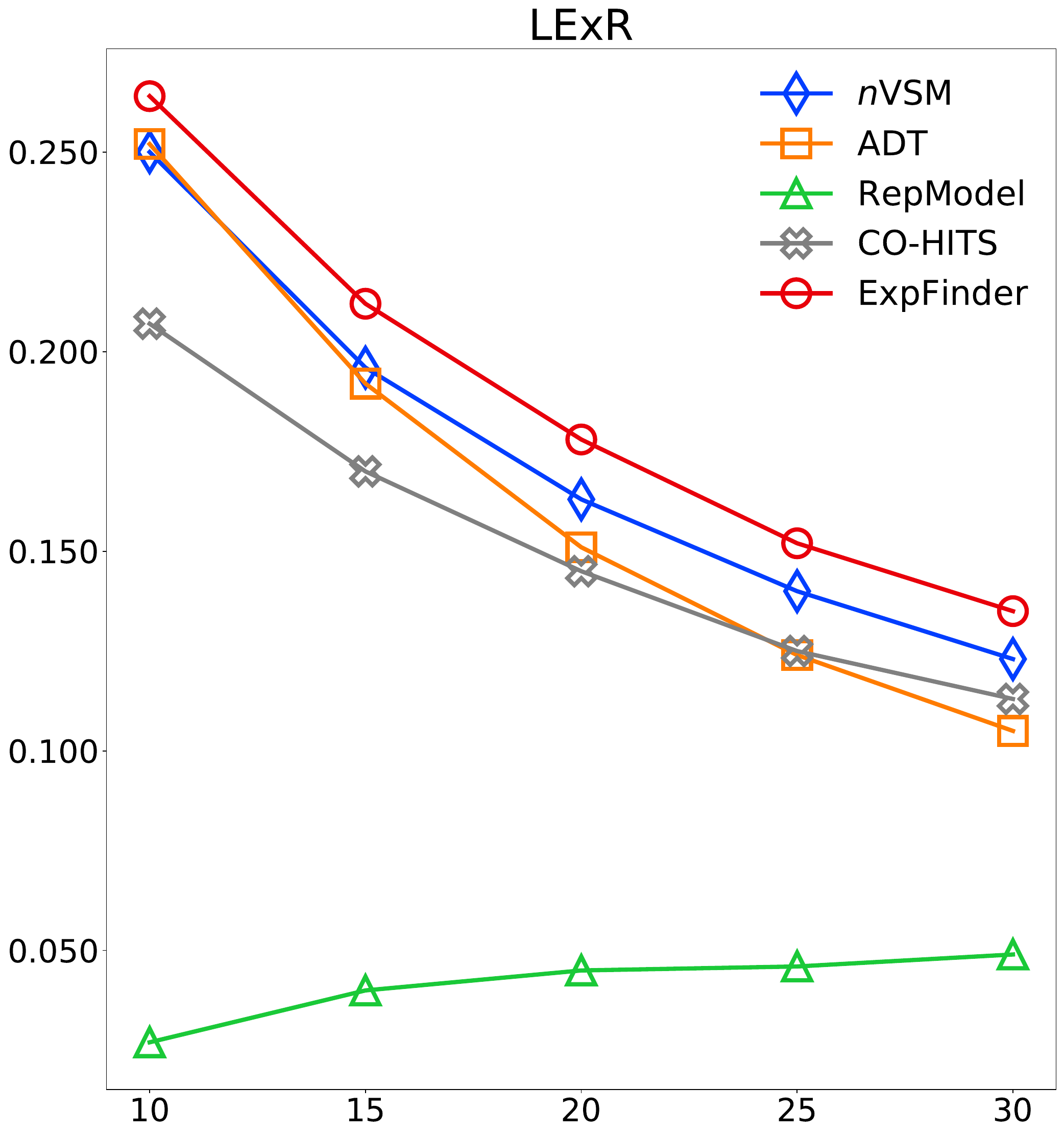} &  \includegraphics[trim=1.8cm 0cm 0cm 0cm,width=39mm,height=40mm]{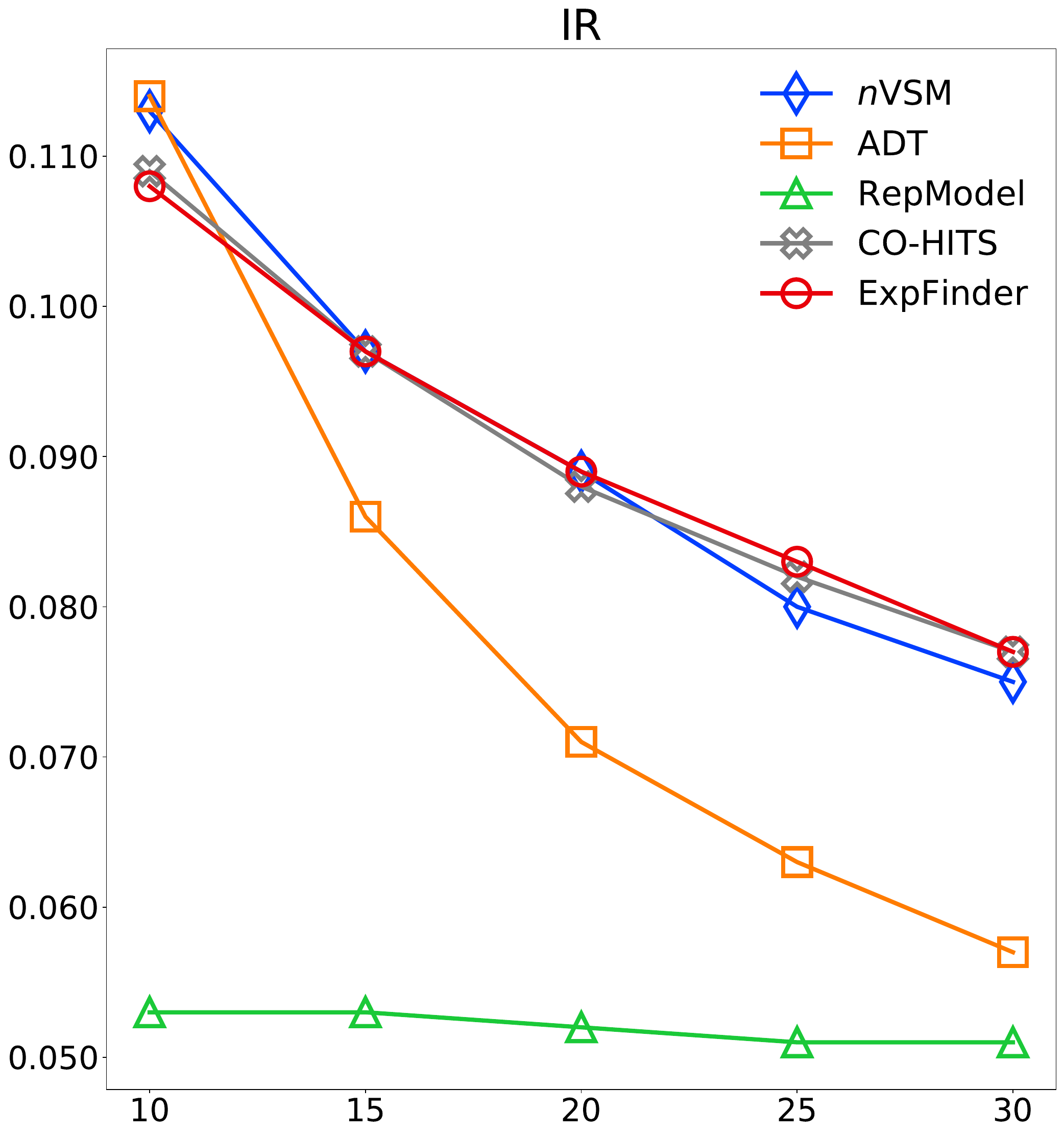} & \includegraphics[trim=1.8cm 0cm 0cm 0cm,width=39mm,height=40mm]{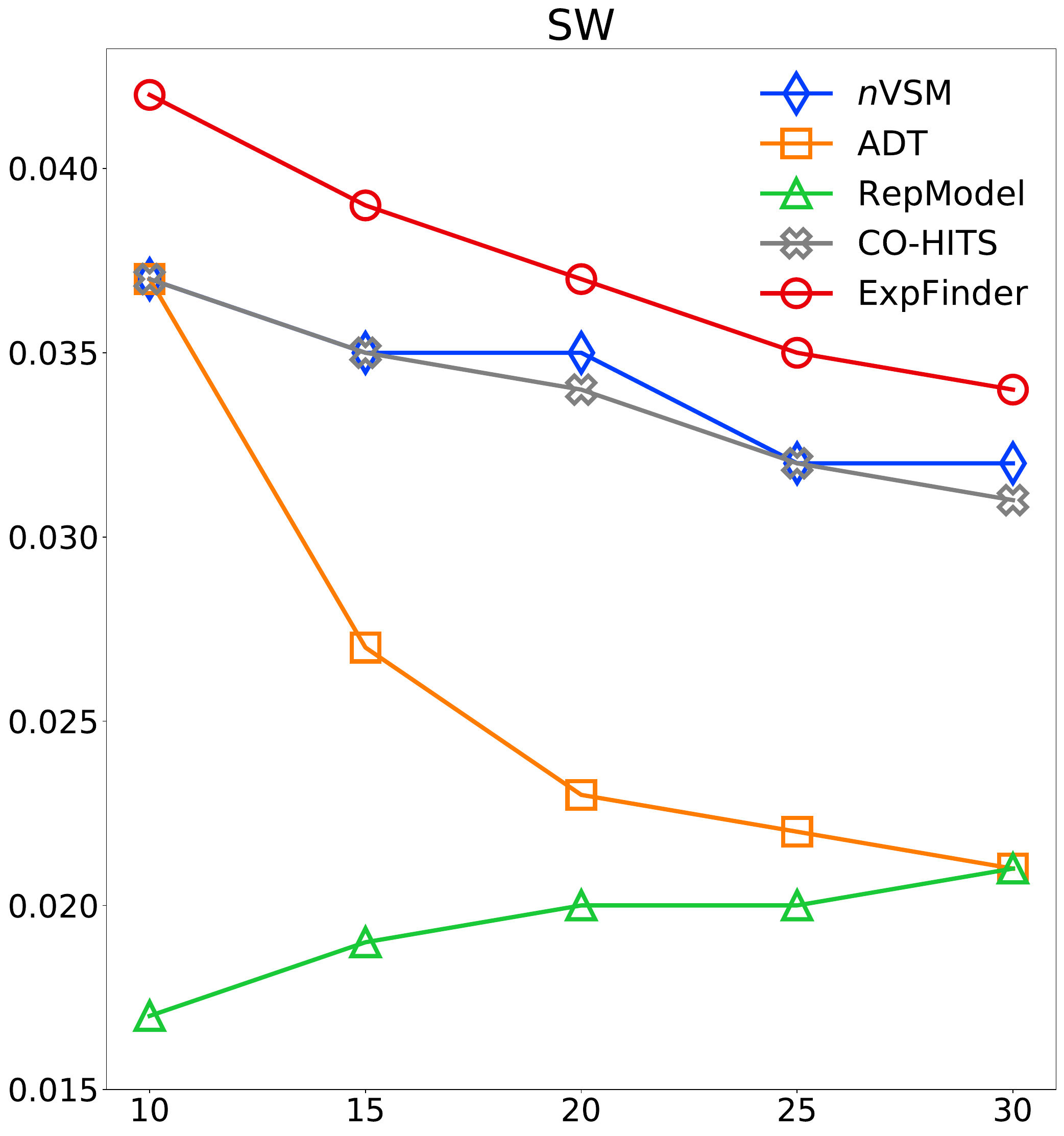} &  \includegraphics[trim=1.8cm 0cm 0cm 0cm,width=39mm,height=40mm]{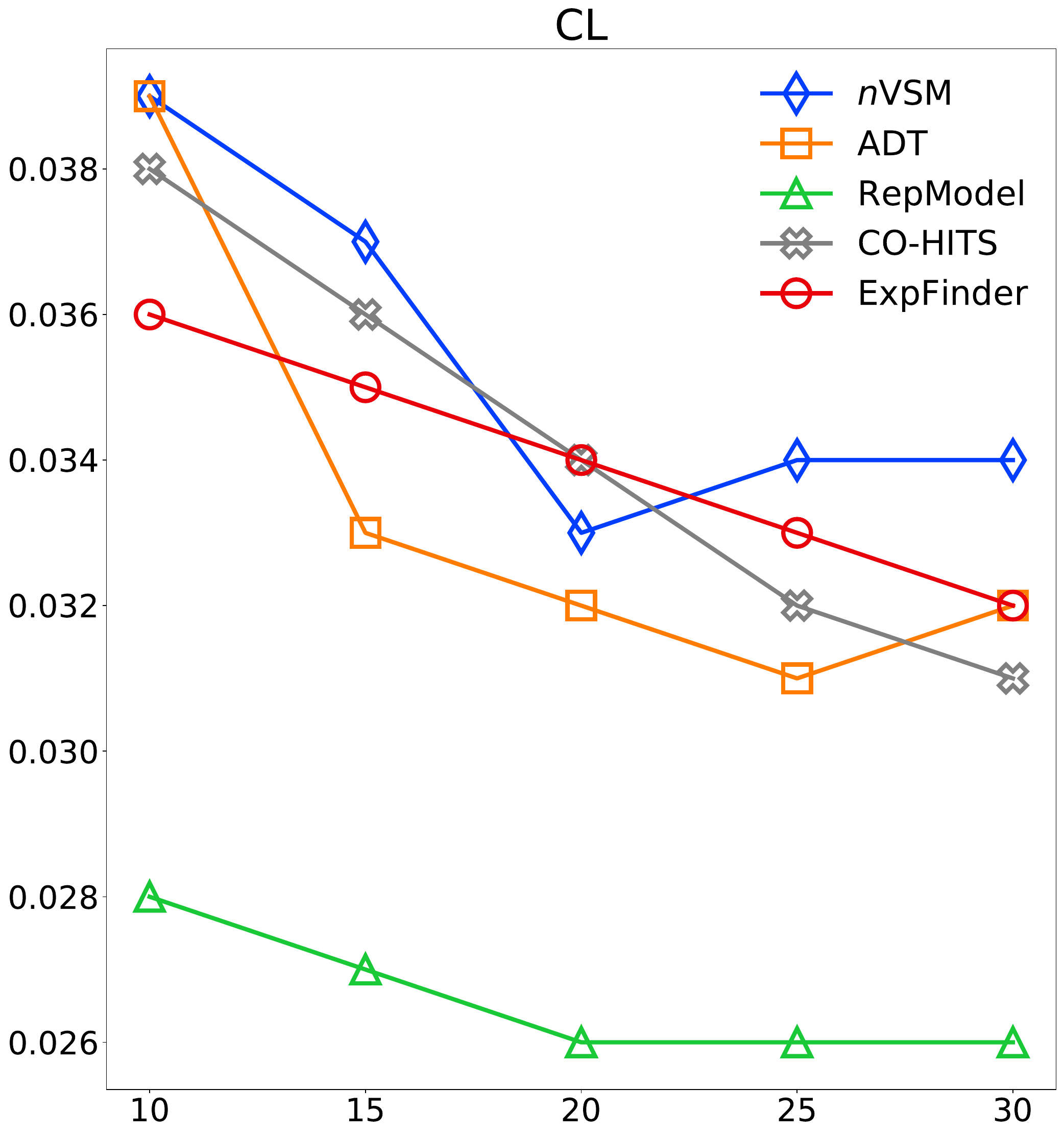}
    \end{tabular}
    \caption{Comparison on the average precision (AP) values: x-axis shows n of P@n and y-axis shows AP values.}
    \label{tab:eval_APF}
\end{center}
\end{figure*}

\captionsetup[table]{labelfont=bf, skip=0pt}
\begin{table*}[!t]
    \caption{MAP and the the improvement ratio of \EF.}
    \begin{center}
        \scalebox{0.9}{
            \setlength{\tabcolsep}{9pt}
            \begin{tabular}{l r l r l r l r l r l}
            \hline
            & LExR & & IR & &  SW & & CL & & Avg. \\
            \hline
            $n$VSM & 0.666 & (12.1\%) & 0.222 & (15.3\%) & 0.106 & (7.6\%) & 0.106 & (4.7\%) & 0.275 & (11.6\%) \\            
            ADT & 0.574 & (30.1\%) & 0.186 & (37.6\%) & 0.077 & (48.1\%) & 0.106 & (4.7\%) & 0.236 & (30.1\%) \\
            RepModel & 0.260 & (187.3\%) & 0.144 & (77.8\%) & 0.061 & (86.9\%) & 0.070 & (58.6\%) & 0.134 & (129.1\%) \\
            CO-HITS & 0.611 & (22.3\%) & 0.228 & (12.3\%) & 0.104 & (9.6\%) & 0.088 & (26.1\%) & 0.258 & (19.0\%) \\
            \EF & \textbf{0.747} & & \textbf{0.256} & & \textbf{0.114} & & \textbf{0.111} &  & \textbf{0.307} & \\        
            \hline
            \end{tabular}
        }
    \end{center}
    \label{tab:eval_MAPF}
\end{table*}

ADT uses an indirect and weighted tripartite (expert-document-topic) graph, where each triplet contains experts, documents and topics. 
Experts are connected to their documents, and also documents are connected to the topics based on their occurrences. The weight of an edge between an expert $x$ and a document $d$ ($w_{xd}$) corresponds to $p(x|d)$ in Eq.~\ref{eq:pw_2}. The weight of an edge between $d$ and a topic $t$ ($w_{dt}$) is modelled as $p(t|d)$ in Eq.~\ref{eq:pw_2}. Recall that we fixed $p(x|d)$ as 1 in Section~\ref{sec:eval_frame}.
As \EF models $p(t|d)$ as $n$TFIDF, we also model $w_{dt}$ as $n$TFIDF in ADT for the fair comparison.
ADT ranks $x$ given a topic $t$ based on the  score function $s(x,t)$ (the higher the more important):
\begin{equation} \label{eq:adt_xt_1}
    s(x, t) = \sum_{d\in \mc{D}_{x}}{w_{xd} \cdot pweight(d,t)},
\end{equation}
where 
    $pweight(d,t)$ = $\sum_{p \in P(d, t)}{\prod_{i}{w(e_i)}}$
where $p$ is a path between $d$ and $t$ comprising of edges such that $p = e_1e_2 \ldots e_n$; $P(d, t)$ is the set of all possible paths between $d$ and $t$; and $w(e_i)$ is the weight of the $i$-th edge in $p$. 
RepModel~\cite{Daniel:2015} was originally designed to estimate the topic-sensitive reputation of an organisation in the context of scientific research projects. This model uses topic-sensitive CO-HITS given a topic, where an organisation is seen as an expert and a project is seen as a document in our work. Thus, using the CO-HITS notations in Eq.~\ref{eq:hits_topic}, RepModel models $A(x;t)^k$ as  $\sum_{w\in t} w_wA(x;w)^k$ and  $H(d;t)^k$ as $\sum_{w\in t} w_wH(d;w)^k$, where $w_w = \frac{1}{|t|}$. As $\lambda_{x}$ and $\lambda_d$, we use 0.85 as used in \cite{Daniel:2015}. In RepModel, the personalised weights $\alpha_{x;w}$ and $\alpha_{d;w}$ are defined as: $\alpha_{d;w}$ = $tf(d,w)$ denoting the term frequency of $w$ divided by $|d|$;
and $\alpha_{x;w}$ = ${s(x,w)}$ if $w$ appears in $\mc{D}_x$, and 0, otherwise. $s(x,w)$ is defined as $1 - \frac{\max_s - weight(x,w)}{\max_s - \min_s }$, where $\max_s$ and $\min_s$ are the max and min values of $weight(x,w)$ for all experts, and $weight(x,w)$ is the weight of $x$ on $w$ calculated by the number of documents in $\mc{D}_x$ where $w$ appears. We also set $k$=5 as done for \hits.

For the fair comparison between \hits and CO-HITS, as \hits, personalised weights $\alpha_{x;t}$ and $\alpha_{d;t}$ in CO-HITS in Eq.~\ref{eq:hits_topic} are set as $\mb{TX}_{i(t),i(x)}$ and $\mb{TD}_{i(t),i(d)}$, respectively. Following the same experiment in Section \ref{sec:lamb_eval}, we found that the best values for $\lambda_{x}$ and $\lambda_{d}$ are chosen as 1.0 and 1.0, respectively, for CO-HITS. 
We also fix $k$ as 5 in Eq.~\ref{eq:hits_topic} as \EF. In our comparison below, CO-HITS indicates an alternative \EF form incorporating $n$VSM into CO-HITS.

\begin{figure*}[!t]
\begin{center}
    \begin{tabular}{llll}
      \includegraphics[trim=1.8cm 0cm 0cm 0cm,width=40mm,height=40mm]{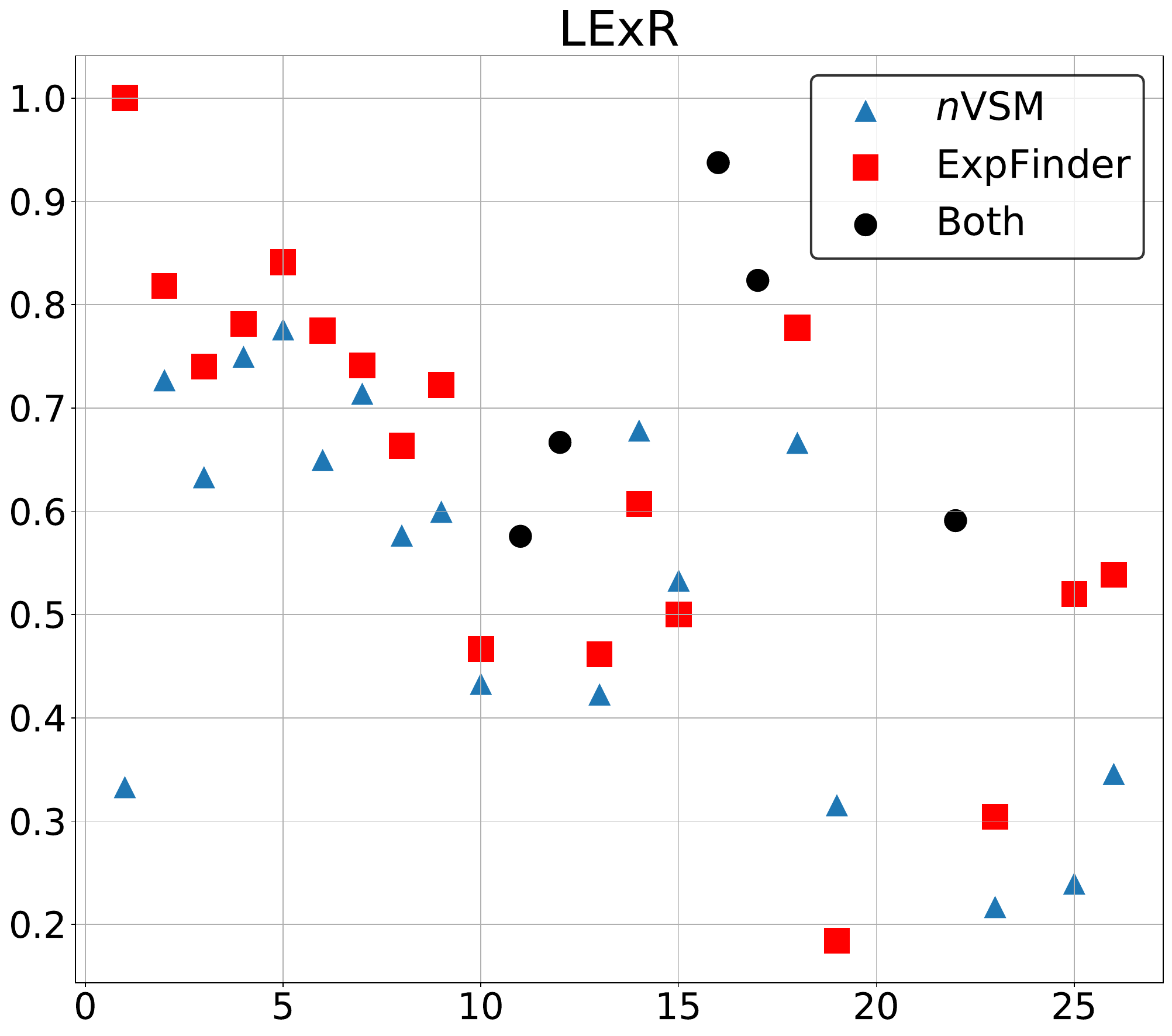} &  \includegraphics[trim=1.8cm 0cm 0cm 0cm,width=40mm,height=40mm]{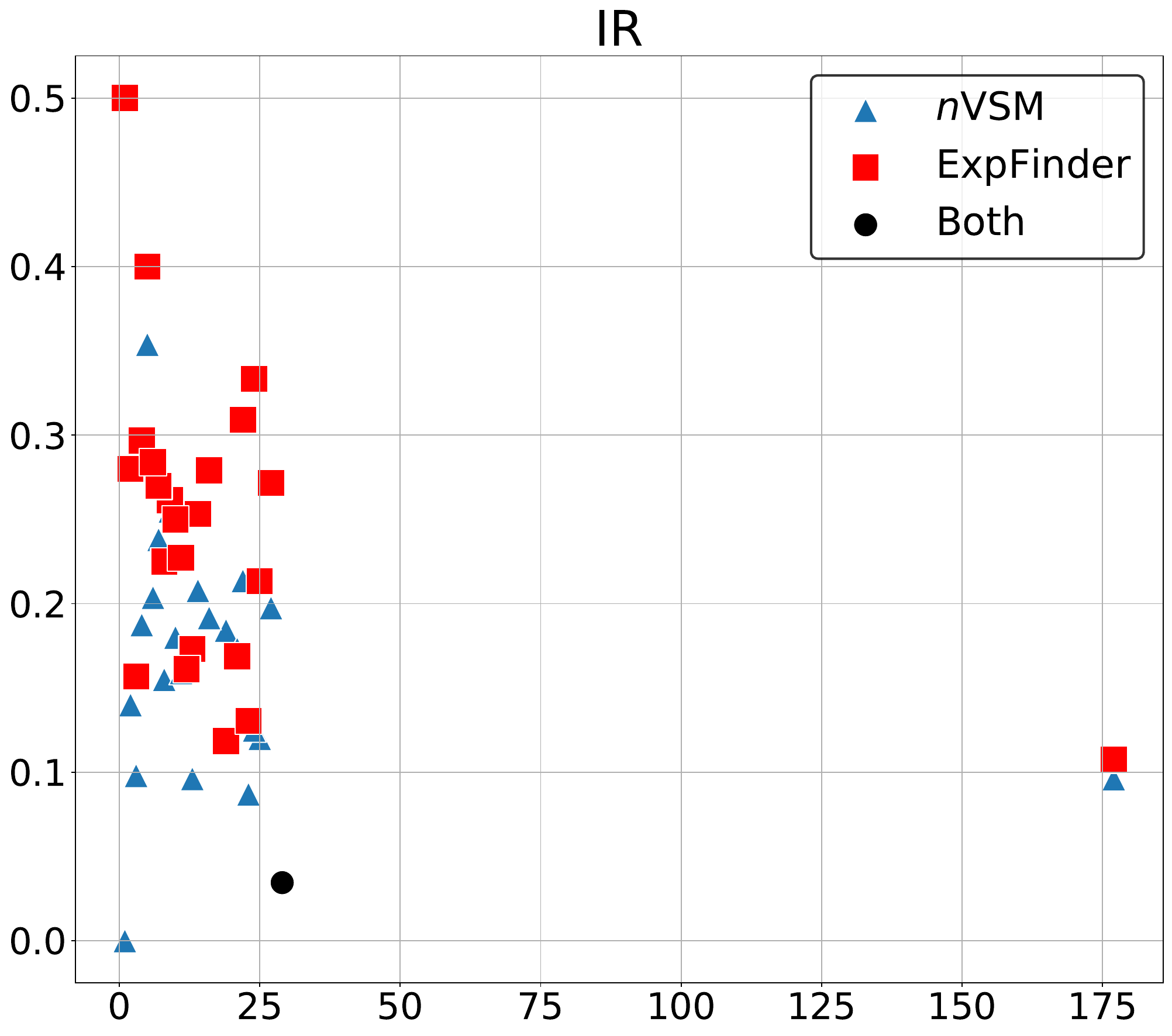} & \includegraphics[trim=1.8cm 0cm 0cm 0cm,width=40mm,height=40mm]{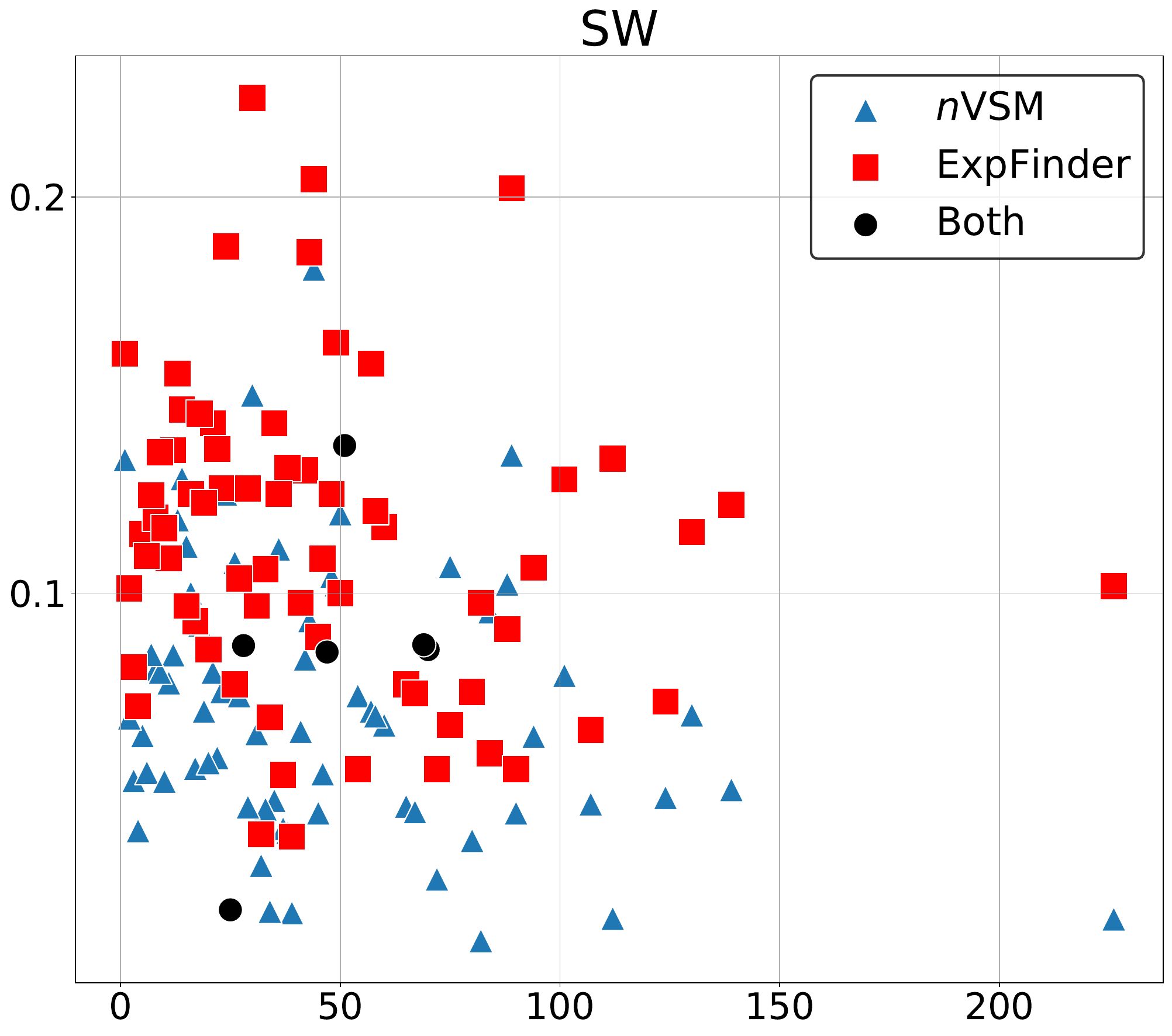} &  \includegraphics[trim=1.8cm 0cm 0cm 0cm,width=40mm,height=40mm]{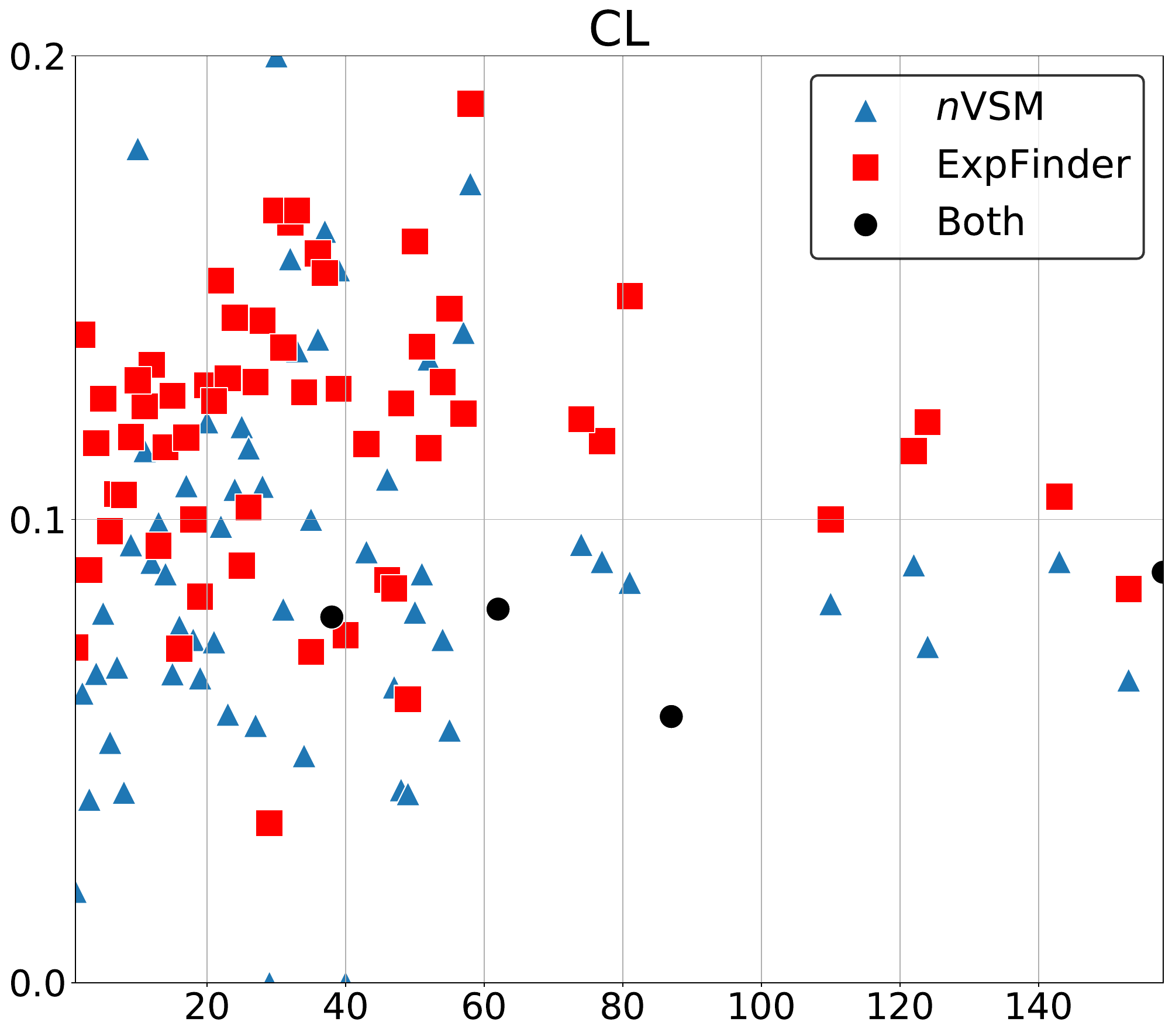}
    \end{tabular}
    \caption{MAP values of $n$VSM and \EF: the x-axis shows topic coverage values, and the y-axis shows the MAP values (topic coverage of a topic $t$ means the number of known experts associated with $t$).}
    \label{fig:eval_ap_expert}
\end{center}
\end{figure*}

Fig. \ref{tab:eval_APF} shows the evaluation results based on the AP values with n (n=10,  15,  $\ldots$, 30) of P@n for all topics. First, when comparing \EF with ADT, although ADT is slightly better than \EF over two datasets IR and CL at n=10, \EF largely outperforms at all values for n of P@n, i.e. n=15, $\ldots$, 30, on all the datasets. It is also clear that \EF is substantially better than RepModel on all x-axis values.
Second, \EF is consistently better than CO-HITS on LExR and SW, and very similar to each other on IR. On CL, CO-HITS is better than at n=10 and n=15, but similar at n=20, and worse than \EF at n=25 and n=30. Overall, these results also show that \hits can have a competitive potential for improving the performance over CO-HITS. Third, to determine the impact of incorporating $n$VSM into \hits in \EF, let us compare \EF with $n$VSM. As observed, \EF is clearly and consistently better than $n$VSM over LExR and SW, although they are similar on IR and $n$VSM looks better than \EF on CL. On average, we observe that our ensemble model \EF combining $n$VSM with \hits is observed to be more powerful than only $n$VSM. 
Table~\ref{tab:eval_MAPF} shows the evaluation results in MAP. The best one is denoted in boldface. As observed, \EF outperforms all the methods in 11.6\%, 30.1\%, 129.1\% and 19.0\% over $n$VSM, ADT, RepModel and CO-HITS, respectively, on average. Interestingly, $n$VSM is observed as the second better one. This also shows that our $n$VSM for expert finding is more competitive than the compared GMs.

Finally, it may be also worth analysing the distribution of the MAP values of a model across topics based on \textit{topic coverage} in each dataset. In our context, the topic coverage of a topic $t$ means the number of known experts having expertise $t$. This enlightens how the model particularly works better or worse at which topic coverage values. Intuitively, it may be harder to find experts for topics whose topic coverage is lower. For this analysis, we pay only attention to our two models $n$VSM and \EF. By comparing their distributions, we can identify which model is better than the other on what topic coverage values. 

The analysis results are seen in Fig.~\ref{fig:eval_ap_expert}. In each plot, each value on the x-axis shows a {topic coverage} value. Each value on the y-axis shows the MAP value of \textit{the set of topics with the same topic coverage}. For example, on LExR, the value 5 on the x-axis means that the topic coverage is 5. The corresponding y-axis value 0.85 indicates that the MAP value of the set of topics with the topic coverage 5 is 0.85. Each MAP value is also calculated using n=30 of P@n. Each black circle represents that both of $n$VSM and \EF have the same MAP value. We can observe the following: (1) On LExR, \EF dominantly outperforms or is equal to $n$VSM over all the topic coverage values except the 4 cases (i.e. 13, 15, 18, and 19); (2) On IR, \EF also shows its improvement over $n$VSM over most of the topic coverage values, while one MAP value is the same on the topic coverage 26; (3) On SW, we observe that \EF prevailingly outperforms $n$VSM across most of the topic values as \EF's distribution is predominantluy higher than $n$VSM's distribution; (4) On CL, although $n$VSM is better than \EF on the topic coverage values 10 and 30, \EF is observed notably better than $n$VSM over the other topic coverage values; and (5) there seems no clear clue that \EF performs particularly better on which range of topic coverage. For example, on LExR, \EF seems to perform better in smaller topic coverage values, as the MAP values from 0 to 5 are clearly higher than those from 6 to 15 on the axis. But this pattern is not consistent with the datasets on SW and CL, where \EF performs better on larger topic coverage values. In conclusion, the above observations may indicate that overall \EF outperforms $n$VSM regardless of topic coverage, showing the validity of the design paradigm of \EF, that is, incorporating $n$VSM into \hits can have powerful capability for expert finding.

\subsection{Discussions and Future Work}\label{sec:discussion}



Using the four datasets from academic domains, we evaluated \EF and its two key components  $n$VSM and \hits, and compared the results with  other expert finding models: TF-IDF based VSM (denoted as TFIDF), DLM-0.5~\cite{Balog:2009} and DLM-0.6~\cite{Wang:2015}, WISER~\cite{WISER:2019}, ADT~\cite{Gollapalli:2013} and RepModel~\citep{Daniel:2015}. 

We showed the capability of $n$VSM using different AP values (n=10, 15, $\ldots$, 30) and MAP, in comparison with TFIDF, DLM-0.5 and DLM-0.6. 
On average, the improvement ratio of $n$VSM over them was turned out as from 21.7\% to 133.1\% in MAP. We also presented the empirical method for finding the best values of the two parameters used in \EF, $\lambda_x$ and $\lambda_d$, based on the ranking of the MAP values. Moreover, we showed how much \EF performs better than all the compared methods in Table~\ref{tab:eval_MAPC} and Table~\ref{tab:eval_MAPF} in MAP. That is, we showed that \EF improves DLM-0.5 and DLM-0.6 in 127.5\% and 160.2\%, respectively; TFIDF in 35.9\%; WISER in 71.5\%; ADT in 31\%; $n$VSM in 11.6\%; RepModel in 129.1\%; and CO-HITS in 19.0\%. 
Further, we showed that \EF incorporating $n$VSM into \hits indeed improves only $n$VSM. It means that exploiting network propagation effects on \ecg using \hits with the outcome of $n$VSM can contribute to better estimating topic-sensitive weights of experts. Also, by comparing \EF with a model combining $n$VSM with CO-HITS, we proved that \hits can be an effective approach for improving CO-HITS for expert finding.
Finally, we analysed that \EF works well regardless of topic coverage values. Our all evaluation results reinforce our motivation of designing \EF that the proposed ensemble model \EF for expert finding is effective and competitive for expert finding.


As future work, it could be worth to try to find a way for improving precision for expert finding. As we have observed in Table~\ref{tab:eval_MAPF}, the average MAP value of \EF is 0.307 across the four datasets. In the literature, we can also observe the similar MAP results. For example, WISER~\cite{WISER:2019} reported that its best MAP values are 0.214 and 0.363 on the two datasets, BMExpert~\cite{Wang:2015} also showed 0.06 as the best MAP value of the DLM~\cite{Balog:2009} on the single dataset, and ADT~\cite{Gollapalli:2013} also showed its best MAP values are 0.0943 and 0.1986 on the two datasets. 
As another future work, we plan to accommodate a general expertise knowledge source as ~\cite{WISER:2019}, e.g. Wikipedia, into \EF to see its potential for enhancing
\EF's capability. Another interesting future work is that we could examine graph embedding techniques for expert finding. One idea would be that we extend \ecg by constructing an expert-document-topic graph based on their semantic relationships. Then we can train a machine to transform nodes, edges and their features into a vector space while maximally preserving their relationship information. Once we would be successfully able to map such a graph to a vector space,  we could estimate the importance of experts given documents or topics by measuring their similarity (or relevance) between experts and documents or between experts and topics in terms of their corresponding vector values.


\section{Conclusion}\label{sec:conclusion}
In this paper, we proposed \EF a novel ensemble model for expert finding. We presented the design of \EF and conducted comprehensive empirical experiments to evaluate and validate its effectiveness using 
four publicly accessible datasets (LExR~\cite{Vitor:2016}, Information Retrieval, Semantic Web and Computational Linguistics in DBLP dataset \citep{bordea2013benchmarking}) from the academic domains.  The novelty  of \EF is in its incorporation of a novel $N$-gram vector space model ($n$VSM) into \hits. The key to designing $n$VSM is to utilise the state-of-the-art IDF method~\citep{nidf:2017} for estimating the topic-sensitive weights of experts given a topic. 
Such estimated weights are further improved by incorporating them into \ecg using \hits. Our novelty of \hits is to design two variation schemes of CO-HITS~\cite{Hongbo:2009}, thus proposing a unified formula for successfully integrating $n$VSM with \hits. We showed comprehensive evaluation, comparing \EF with the six representative models, that is,  TF-IDF based vector space models (TFIDF), two document language models (DLM) \cite{Balog:2009, WISER:2019},  two graph-based models (GM)~\citep{Gollapalli:2013,Daniel:2015}, and topic-sensitive CO-HITS~\cite{Hongbo:2009}. We showed  \EF is a highly effective ensemble model for expert finding, outperforming TFIDF with 35.9\%, DLMs with 71.5\% - 160.2\%, GMs with 31\% - 129.1\%, and topic-sensitive CO-HITS with 19.0\%. 



\ifCLASSOPTIONcaptionsoff
  \newpage
\fi

\footnotesize{{\bibliographystyle{IEEEtran}}}
\bibliography{ybk}

\begin{thebibliography}{10}
\providecommand{\url}[1]{#1}
\csname url@samestyle\endcsname
\providecommand{\newblock}{\relax}
\providecommand{\bibinfo}[2]{#2}
\providecommand{\BIBentrySTDinterwordspacing}{\spaceskip=0pt\relax}
\providecommand{\BIBentryALTinterwordstretchfactor}{4}
\providecommand{\BIBentryALTinterwordspacing}{\spaceskip=\fontdimen2\font plus
\BIBentryALTinterwordstretchfactor\fontdimen3\font minus
  \fontdimen4\font\relax}
\providecommand{\BIBforeignlanguage}[2]{{%
\expandafter\ifx\csname l@#1\endcsname\relax
\typeout{** WARNING: IEEEtran.bst: No hyphenation pattern has been}%
\typeout{** loaded for the language `#1'. Using the pattern for}%
\typeout{** the default language instead.}%
\else
\language=\csname l@#1\endcsname
\fi
#2}}
\providecommand{\BIBdecl}{\relax}
\BIBdecl

\bibitem{husain2019expert}
O.~Husain, N.~Salim, R.~A. Alias, S.~Abdelsalam, and A.~Hassan, ``Expert
  finding systems: A systematic review,'' \emph{Applied Sciences}, vol.~9,
  no.~20, p. 4250, 2019.

\bibitem{Stankovic2010LookingFE}
M.~Stankovic, C.~Wagner, J.~Jovanovic, and P.~Laublet, ``Looking for experts?
  what can linked data do for you?'' in \emph{Proceedings of the WWW2010
  Workshop on Linked Data on the Web}, 2010.

\bibitem{gonccalves2019automated}
R.~Gon{\c{c}}alves and C.~F. Dorneles, ``Automated expertise retrieval: A
  taxonomy-based survey and open issues,'' \emph{ACM Computing Surveys (CSUR)},
  vol.~52, no.~5, pp. 1--30, 2019.

\bibitem{chuang2014combining}
C.~T. Chuang, K.~H. Yang, Y.~L. Lin, and J.~H. Wang, ``Combining query terms
  extension and weight correlative for expert finding,'' in \emph{2014
  IEEE/WIC/ACM International Joint Conferences on Web Intelligence (WI) and
  Intelligent Agent Technologies (IAT)}, vol.~1.\hskip 1em plus 0.5em minus
  0.4em\relax IEEE, 2014, pp. 323--326.

\bibitem{alhabashneh2017fuzzy}
O.~Alhabashneh, R.~Iqbal, F.~Doctor, and A.~James, ``Fuzzy rule based profiling
  approach for enterprise information seeking and retrieval,''
  \emph{Information Sciences}, vol. 394, pp. 18--37, 2017.

\bibitem{Balog:2009}
K.~Balog, L.~Azzopardi, and M.~de~Rijke, ``A language modeling framework for
  expert finding,'' \emph{Information Processing \& Management}, vol.~45,
  no.~1, pp. 1 -- 19, 2009.

\bibitem{Wang:2015}
B.~Wang, X.~Chen, H.~Mamitsuka, and S.~Zhu, ``Bmexpert\: Mining medline for
  finding experts in biomedical domains based on language model,''
  \emph{IEEE/ACM Trans. Comput. Biol. Bioinformatics}, vol.~12, no.~6, pp.
  1286--1294, Nov. 2015.

\bibitem{WISER:2019}
P.~Cifariello, P.~Ferragina, and M.~Ponza, ``{WISER: A semantic approach for
  expert finding in academia based on entity linking},'' \emph{Information
  Systems}, vol.~82, pp. 1 -- 16, 2019.

\bibitem{Gollapalli:2013}
S.~D. Gollapalli, P.~Mitra, and C.~L. Giles, ``Ranking experts using
  author-document-topic graphs,'' in \emph{Proceedings of the 13th ACM/IEEE-CS
  Joint Conference on Digital Libraries}, ser. JCDL '13, 2013, pp. 87--96.

\bibitem{Campbell:2003}
C.~S. Campbell, P.~P. Maglio, A.~Cozzi, and B.~Dom, ``Expertise identification
  using email communications,'' in \emph{Proceedings of the Twelfth
  International Conference on Information and Knowledge Management}, ser. CIKM
  '03, 2003, p. 528–531.

\bibitem{Yeniterzi:2014}
R.~Yeniterzi and J.~Callan, ``Constructing effective and efficient
  topic-specific authority networks for expert finding in social media,'' 2014,
  p. 45–50.

\bibitem{faisal2019expert}
M.~S. Faisal, A.~Daud, A.~U. Akram, R.~A. Abbasi, N.~R. Aljohani, and
  I.~Mehmood, ``Expert ranking techniques for online rated forums,''
  \emph{Computers in Human Behavior}, vol. 100, pp. 168--176, 2019.

\bibitem{bok2019expert}
K.~Bok, I.~Jeon, J.~Lim, and J.~Yoo, ``Expert finding considering dynamic
  profiles and trust in social networks,'' \emph{Electronics}, vol.~8, no.~10,
  p. 1165, 2019.

\bibitem{Sziklai:2018}
B.~Sziklai, ``How to identify experts in a community?'' \emph{International
  Journal of Game Theory}, vol.~47, no.~1, pp. 155--173, 2018.

\bibitem{nidf:2017}
M.~Shirakawa, T.~Hara, and S.~Nishio, ``{IDF for Word N-Grams},'' \emph{ACM
  Trans. Inf. Syst.}, vol.~36, no.~1, Jun. 2017.

\bibitem{Hongbo:2009}
H.~Deng, M.~R. Lyu, and I.~King, ``{A Generalized CO-HITS Algorithm and Its
  Application to Bipartite Graphs},'' in \emph{Proceedings of the 15th ACM
  SIGKDD International Conference on Knowledge Discovery and Data Mining},
  2009, p. 239–248.

\bibitem{Vitor:2016}
V.~Mangaravite, R.~L. Santos, I.~S. Ribeiro, M.~A. Gon\c{c}alves, and A.~H.
  Laender, ``{The LExR Collection for Expertise Retrieval in Academia},'' in
  \emph{Proceedings of the 39th International ACM SIGIR Conference on Research
  and Development in Information Retrieval}, 2016, p. 721–724.

\bibitem{bordea2013benchmarking}
G.~Bordea, T.~Bogers, and P.~Buitelaar, ``Benchmarking domain-specific expert
  search using workshop program committees,'' in \emph{Proceedings of the 2013
  workshop on Computational scientometrics: theory \& applications}, 2013, pp.
  19--24.

\bibitem{Daniel:2015}
D.~Schall, \emph{A Social Network-Based Recommender Systems}.\hskip 1em plus
  0.5em minus 0.4em\relax Springer, 2015.

\bibitem{riahi2012finding}
F.~Riahi, Z.~Zolaktaf, M.~Shafiei, and E.~Milios, ``Finding expert users in
  community question answering,'' in \emph{Proceedings of the 21st
  International Conference on World Wide Web}, 2012, pp. 791--798.

\bibitem{Schunk:2010}
L.~K. Schunk and G.~Cong, ``Using transactional data from erp systems for
  expert finding,'' in \emph{Database and Expert Systems Applications}, 2010,
  pp. 267--276.

\bibitem{Mangaravite:2016}
V.~Mangaravite and R.~L. Santos, ``On information-theoretic document-person
  associations for expert search in academia,'' in \emph{Proceedings of the
  39th International ACM SIGIR Conference on Research and Development in
  Information Retrieval}, ser. SIGIR '16.\hskip 1em plus 0.5em minus
  0.4em\relax Association for Computing Machinery, 2016, p. 925–928.

\bibitem{van2016unsupervised}
C.~Van~Gysel, M.~de~Rijke, and M.~Worring, ``Unsupervised, efficient and
  semantic expertise retrieval,'' in \emph{Proceedings of the 25th
  International Conference on World Wide Web}, 2016, pp. 1069--1079.

\bibitem{BM25:2009}
S.~Robertson and H.~Zaragoza, ``The probabilistic relevance framework: Bm25 and
  beyond,'' \emph{Foundations and Trends® in Information Retrieval}, vol.~3,
  no.~4, pp. 333--389, 2009.

\bibitem{jiang2016exploiting}
X.~Jiang, X.~Sun, Z.~Yang, H.~Zhuge, and J.~Yao, ``Exploiting heterogeneous
  scientific literature networks to combat ranking bias: Evidence from the
  computational linguistics area,'' \emph{Journal of the Association for
  Information Science and Technology}, vol.~67, no.~7, pp. 1679--1702, 2016.

\bibitem{koumenides2014ranking}
C.~L. Koumenides and N.~R. Shadbolt, ``Ranking methods for entity-oriented
  semantic web search,'' \emph{Journal of the Association for Information
  Science and Technology}, vol.~65, no.~6, pp. 1091--1106, 2014.

\bibitem{zhang2007expertise}
J.~Zhang, M.~S. Ackerman, and L.~Adamic, ``Expertise networks in online
  communities: structure and algorithms,'' in \emph{Proceedings of the 16th
  international conference on World Wide Web}, 2007, pp. 221--230.

\bibitem{wang2013expertrank}
G.~A. Wang, J.~Jiao, A.~S. Abrahams, W.~Fan, and Z.~Zhang, ``Expertrank: A
  topic-aware expert finding algorithm for online knowledge communities,''
  \emph{Decision support systems}, vol.~54, no.~3, pp. 1442--1451, 2013.

\bibitem{kundu2019formulation}
D.~Kundu and D.~P. Mandal, ``Formulation of a hybrid expertise retrieval system
  in community question answering services,'' \emph{Applied Intelligence},
  vol.~49, no.~2, pp. 463--477, 2019.

\bibitem{deng2011enhanced}
H.~Deng, I.~King, and M.~R. Lyu, ``Enhanced models for expertise retrieval
  using community-aware strategies,'' \emph{IEEE Transactions on Systems, Man,
  and Cybernetics, Part B (Cybernetics)}, vol.~42, no.~1, pp. 93--106, 2011.

\bibitem{liu2013integrating}
D.-R. Liu, Y.-H. Chen, W.-C. Kao, and H.-W. Wang, ``Integrating expert profile,
  reputation and link analysis for expert finding in question-answering
  websites,'' \emph{Information processing \& management}, vol.~49, no.~1, pp.
  312--329, 2013.

\bibitem{Aslay:2013}
c.~Aslay, N.~O'Hare, L.~M. Aiello, and A.~Jaimes, ``Competition-based networks
  for expert finding,'' in \emph{Proceedings of the 36th International ACM
  SIGIR Conference on Research and Development in Information Retrieval}, ser.
  SIGIR '13, 2013, p. 1033–1036.

\bibitem{Mahani:2018}
N.~Torkzadeh Mahani, M.~Dehghani, M.~S. Mirian, A.~Shakery, and K.~Taheri,
  ``Expert finding by the dempster-shafer theory for evidence combination,''
  \emph{Expert Systems}, vol.~35, no.~1, 2018.

\bibitem{Kang:2014}
Y.-B. Kang, P.~Delir~Haghighi, and F.~Burstein, ``Cfinder: An intelligent key
  concept finder from text for ontology development,'' \emph{Expert Syst.
  Appl.}, vol.~41, no.~9, p. 4494–4504, 2014.

\bibitem{Roelleke:08}
T.~Roelleke and J.~Wang, ``Tf-idf uncovered: A study of theories and
  probabilities,'' in \emph{Proceedings of the 31st Annual International ACM
  SIGIR Conference on Research and Development in Information Retrieval}, 2008,
  p. 435–442.

\bibitem{lu2013insight}
K.~Lu, ``An insight into vector space modeling and language modeling,'' in
  \emph{iConference}, 2013.

\bibitem{Kleinberg:1999}
J.~M. Kleinberg, ``Authoritative sources in a hyperlinked environment,''
  \emph{J. ACM}, vol.~46, no.~5, p. 604–632, Sep. 1999.

\bibitem{Deng:2008}
H.~Deng, I.~King, and M.~R. Lyu, ``{Formal Models for Expert Finding on DBLP
  Bibliography Data},'' in \emph{Proceedings of the 2008 Eighth IEEE
  International Conference on Data Mining}, ser. ICDM '08, 2008, pp. 163--172.

\bibitem{Beltagy2019SciBERT}
I.~Beltagy, K.~Lo, and A.~Cohan, ``Scibert: Pretrained language model for
  scientific text,'' in \emph{EMNLP}, 2019.

\bibitem{Macdonald:2006}
C.~Macdonald and I.~Ounis, ``Voting for candidates: Adapting data fusion
  techniques for an expert search task,'' in \emph{Proceedings of the 15th ACM
  International Conference on Information and Knowledge Management}, ser. CIKM
  '06, 2006, p. 387–396.

\bibitem{Yuanhua:2011}
Y.~Lv and C.~Zhai, ``Adaptive term frequency normalization for bm25,'' in
  \emph{Proceedings of the 20th ACM International Conference on Information and
  Knowledge Management}, ser. CIKM '11.\hskip 1em plus 0.5em minus 0.4em\relax
  Association for Computing Machinery, 2011, p. 1985–1988.

\end{thebibliography}

%

\begin{IEEEbiography}[{\includegraphics[width=1in,height=1.25in,clip,keepaspectratio]{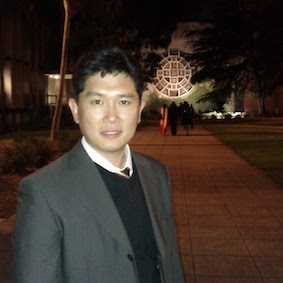}}]
{Yong-Bin Kang} received his PhD in Information Technology from Monash University in 2011. Currently, he
is a senior data science research fellow in the fields of natural language processing, machine learning and data mining at Swinburne University of Technology. He has been working on large industrial, multi-disciplinary research projects such as patent analytics, clinical data analytics, scientific-article analytics, social-media data analytics, expert finding and matching, and machine learning algorithms and applications. 
\end{IEEEbiography}
\vspace{-30pt}

\begin{IEEEbiography}[{\includegraphics[width=1in,height=1in,clip,keepaspectratio]{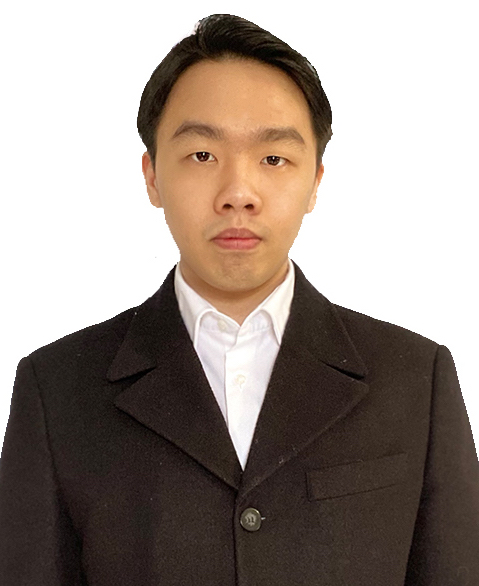}}]{Hung Du} is currently studying his final year Bachelor degree in Computer Science  at Swinburne University of Technology, Australia and working as research assistant. His research interest include Natural Language Processing, Reinforcement Learning, Data Mining and Applied Machine Learning.
\end{IEEEbiography}
\vspace{-30pt}

\begin{IEEEbiography}[{\includegraphics[width=1in,height=1.25in,clip,keepaspectratio]{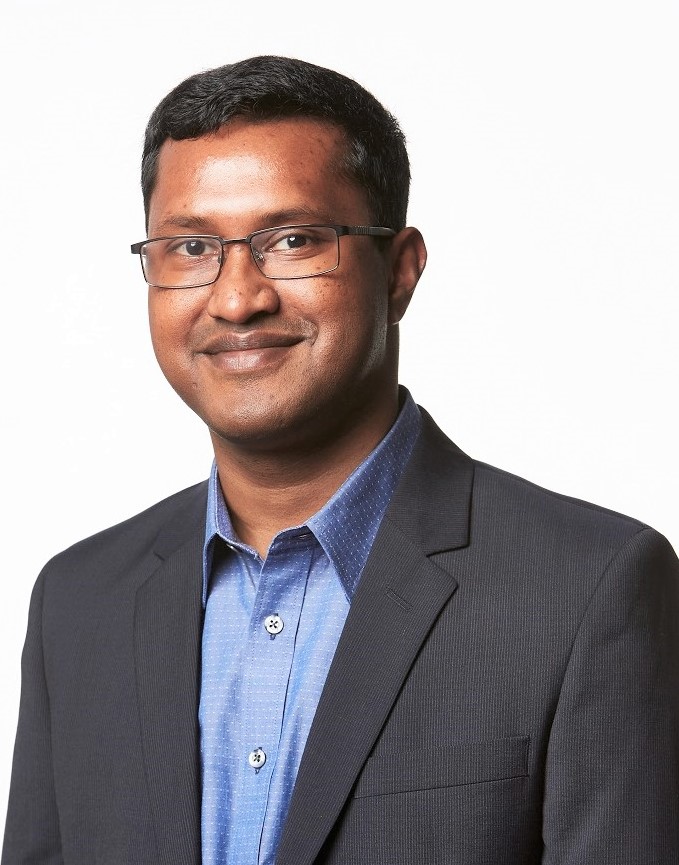}}]{Abdur Rahim Mohammad Forkan}  received his PhD degree in Computer Science from the RMIT University, Australia in 2016. He is a Senior Research Engineer at the Digital Innovation Lab, Swinburne University of Technology, Australia and working on several multi-disciplinary industrial research projects. His research interest include Data Analytics, Health Informatics, Pervasive Systems, Cloud Computing, Data Mining and Applied Machine Learning.
\end{IEEEbiography}
\vspace{-30pt}

\begin{IEEEbiography}[{\includegraphics[width=1in,height=1.25in,clip,keepaspectratio]{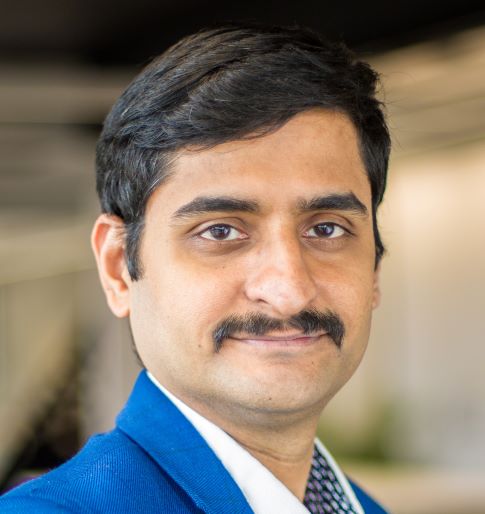}}]{Prem Prakash Jayaraman}
is an associate Professor and the Head of Digital Innovation Lab at Swinburne. 
He is broadly interested in the research areas of Internet of Things, Mobile and Cloud Computing, Health Informatics and application of Data Science techniques and methodologies in real-world settings. He was a key contributor and one of the architects of the Open Source Internet of Things project (OpenIoT) that has won the prestigious Black Duck Rookie of the Year Award in 2013 (https://github.com/OpenIotOrg/openiot). 
\end{IEEEbiography}
\vspace{-30pt}

\begin{IEEEbiography}[{\includegraphics[width=1in,height=1.25in,clip,keepaspectratio]{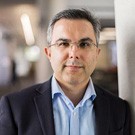}}]{Amir Aryani}
is the Head of the Social Data Analytics (SoDA) Lab at the Swinburne University of Technology. He has experience with large-scale and cross-institution projects in Australia and Europe. His track records include collaboration with high-profile international institutions such as British Library, ORCID, Data Archiving and Network Analysis (DANS, Netherlands),
and funders including ARC, NHMRC, and NIH. He has published high impact journals such as Nature Scientific Data, Metadata and Semantics Research, and Frontiers in Artificial Intelligence and Applications.
\end{IEEEbiography}
\vspace{-30pt}

\begin{IEEEbiography}[{\includegraphics[width=1.0in,height=1.25in,clip,keepaspectratio]{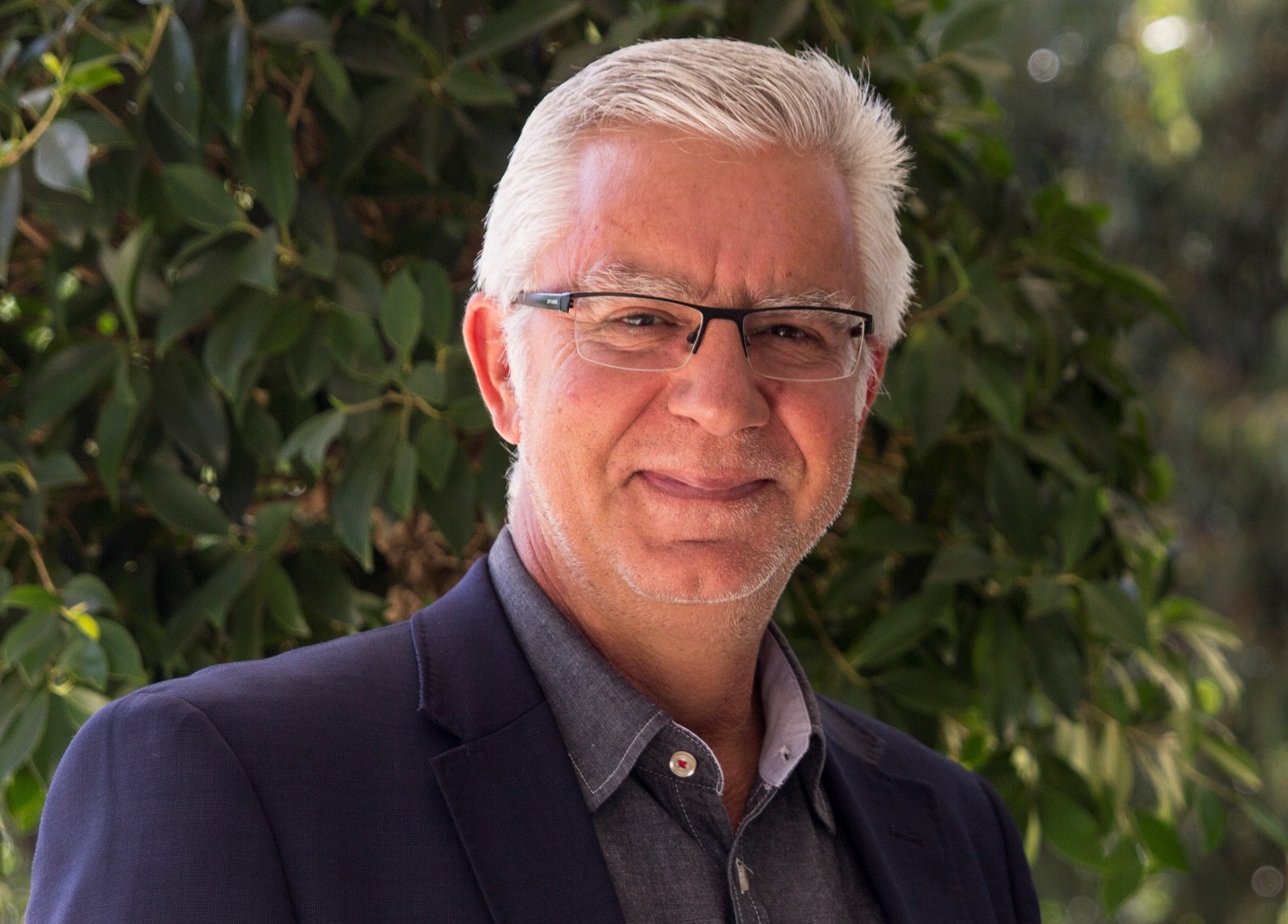}}]{Timos Sellis}
received the PhD degree in Computer Science from the University of California, Berkeley, in 1986. He is now a visiting scientist at Facebook and also adjunct professor at Swinburne. Between 2013 and 2015, he was a professor at RMIT University, Australia, and before 2013, the director of the Institute for the Management of Information Systems and a professor at the National Technical University of Athens, Greece. 
He is a fellow of the IEEE and ACM. 
In 2018, he was awarded the IEEE TCDE Impact Award, in recognition of his impact in the field and for contributions to database systems and data engineering research.
\end{IEEEbiography}
\end{document}